# 4. Electric scanning probe imaging and modification of ferroelectric surfaces

Sergei V. Kalinin[*] and Dawn A. Bonnell

University of Pennsylvania, Philadelphia, PA 19104

[*] Condensed Matter Sciences Division, Oak Ridge National Laboratory, Oak Ridge, TN 37831

## 4.1. SPM Imaging and Control of Ferroelectric Materials

Recent progress in oxide electronic devices including microelectromechanical systems (MEMS), non-volatile ferroelectric memories (FeRAMs), and ferroelectric heterostructures necessitates an understanding of local ferroelectric properties on the nanometer level. This has motivated a number of studies of ferroelectric materials with various scanning probe microscopies (SPM),[1,2,3] many examples of which can be found in this text. The natures of the probe and contrast formation mechanisms in these techniques are vastly different; therefore, SPM images reflect different properties of ferroelectric surfaces. Table I summarizes some of the most common SPM imaging techniques used for the characterization of ferroelectric materials and briefly presents the information obtained.

**Table 1.** SPM Techniques for ferroelectric imaging

| Technique | Measured signal | Relation to ferroelectric properties |
|---|---|---|
| EFM, SSPM | Electrostatic force gradient (EFM), effective surface potential (SSPM) | Characterize electrostatic stray fields above ferroelectric surface induced by surface polarization charge. Sensitive only to out-of–plane polarization component. |
| PFM | Vertical (v-PFM) and lateral (l-PFM) surface displacement induced by tip bias. | Characterizes piezoelectric properties of the surface. Vertical and lateral components of signal are related to in-plane and out-of–plane polarization components. |
| SCM | Voltage derivative of | Based on polarization-induced hysteresis |

|      | tip-surface capacitance | in tip-surface capacitance. Only out-of–plane polarization component can be determined. |
|------|-------------------------|---------------------------------------------------------------------------------------|
| NSOM | Near-field optical properties of the surface | Optical indicatrix of the surface reflects polarization induced anisotropy. Both in-plane and out-of-plane polarization components can be determined. |
| FFM  | Friction forces         | Characterizes effect of polarization charge on surface friction. |
| SNDM | Non-linear dielectric permittivity | Both in-plane and out-of-plane polarization components can be measured. |

On well-defined ferroelectric surfaces, the polarization related surface properties are uniform within the domain and change abruptly at domain walls, providing readily interpretable contrast in SPM measurements. In addition, most SPM techniques allow local poling of ferroelectric materials with subsequent imaging of induced changes. These two factors contribute to the general interest of the SPM community to ferroelectric materials and vice versa. From the materials scientist point of view, the morphological information on domain structure and orientation obtained from SPM images is sufficient for many applications, and numerous observations of local domain dynamics as related to polarization switching processes,[4,5,6] ferroelectric fatigue,[7,8,9,10] phase transitions,[11,12,13,14] mechanical stresses,[15] etc. have been made. However, the true potential of spatially resolved SPM techniques lies in the possibility of *quantitative* measurement of local materials properties.

Quantitative understanding of SPM data is particularly important since most SPM techniques listed in Table I allow spectroscopic measurements, in which the local response is measured as a function of an a external parameter, such as bias, time or temperature. The most widely used is voltage spectroscopy, i.e. local hysteresis loop measurements. However, unlike imaging applications, interpretation of spectroscopic measurements presents a significant challenge. The image formation mechanism in SPM is usually complex and depends sensitively on the details of probe-surface interactions. The interaction volume of the SPM probe is small and minute contamination or damage of the surface precludes imaging. Non-local contribution to the signal due to the non-local part of the probe can be large and comparable to property variations between the domains. This effect usually exhibits as a constant offset that can be ignored in imaging. However, analysis and interpretation of local ferroelectric behavior including variable temperature experiments, phase transitions, hysteresis loop measurements,[16] stress and size effects,[17,18,19] requires quantitative interpretation of the SPM interaction.

This chapter will review two of the most common imaging SPM techniques for ferroelectric imaging. Non-contact electrostatic scanning probe techniques such as Electrostatic Force Microscopy (EFM) and Scanning Surface Potential Micros-copy (SSPM) are used to image electric fields associated with polarization charge on ferroelectric surfaces.[20,21,22,23,24] In ambient, polarization charge is screened by surface electronic states or adsorption which minimize the depolarization field above the surface.[25] Thus non-contact SPMs allow qualitative and quantitative characterization of polarization screening processes related to the polarization-dependent physical and chemical properties of ferroelectric surface. In particular, dynamic behavior of polarization and screening charge can be visualized and quantified *in situ* by variable temperature SPM measurements.

In contrast to non-contact electrostatic SPMs, techniques such as Piezoresponse Force Microscopy, Scanning Capacitance Microscopy and Scanning Near Field Optical Microscopy probe a finite volume of material directly below the probe, i.e. address the subsurface properties of material. Among these techniques, the most widely used currently is Piezoresponse Force Microscopy. In this Chapter, the contrast formation mechanism is analyzed and relative magnitudes of electrostatic vs. electromechanical contributions to PFM interaction for the model case of $c^+$, $c^-$ domains in tetragonal perovskite ferroelectrics are determined. Contrast Mechanism Maps were constructed to delineate the regions with dominant electrostatic and electromechanical interactions depending on experimental conditions. The information accessible by non-contact and contact SPM is complementary and an approach for simultaneous acquisition of SSPM and PFM data is presented.

## 4.2. Non-contact electrostatic imaging of ferroelectric surfaces

### 4.2.1. Electrostatic Imaging Techniques

Electrostatic SPMs are an example of local force-based probes originally introduced by Binnig, Quate and Gerber.[26] Initially, Atomic Force Microscopy (AFM) was designed to measure the strong short-range repulsive forces between a tip and surface. It was almost immediately realized that AFM can be extended to map long range electromagnetic forces.[27,28]

Among the electrostatic force sensitive SPM techniques the most developed are Electrostatic Force Microscopy and Scanning Surface Potential Microscopy (Kelvin Probe Force Microscopy). Both EFM and SSPM are based on the dual pass scheme. The grounded tip first acquires the surface topography using standard intermittent contact AFM. Electrostatic data are collected in the second scan, during which the tip retraces the topographic profile separated from the surface 50 to 100 nm, thereby maintaining a constant tip-sample separation. In EFM, the cantilever is driven mechanically, and the electrostatic force, *F*, between the dc biased conductive tip and the surface results in a change of the cantilever resonant frequency that is proportional to the force gradient[29]

$$\Delta\omega = \frac{\omega_0}{2k}\frac{\mathrm{d}F(z)}{\mathrm{d}z}, \tag{1}$$

where $k$ is the spring constant and $\omega_0$ is the resonant frequency of the cantilever. Resonance is maintained by adjusting the driving frequency $\omega_p$ and the frequency shift $\Delta\omega$ is collected as the EFM image.

In SSPM the cantilever is not driven mechanically; rather, the tip is biased directly by $V_{tip} = V_{dc}+V_{ac}\cos(\omega t)$, where $V_{ac}$ is referred to as the driving voltage. The capacitive force, $F_{cap}(z)$, between the tip and a surface at potential $V_s$ is:

$$F_{cap}(z) = \frac{1}{2}(V_{tip} - V_s)^2 \frac{\partial C(z)}{\partial z}, \tag{2}$$

where $C(z)$ is the tip-surface capacitance dependent on tip geometry, surface topography and tip surface separation, $z$. The first harmonic of the force is

$$F_{1\omega}^{cap}(z) = \frac{\partial C(z)}{\partial z}(V_{dc} - V_s)V_{ac} \tag{3}$$

and feedback is used to nullify this term by adjusting the constant component of the tip bias, $V_{dc}$. This condition is met when $V_{dc}$ is equal to surface potential (defined as $V_{el}$ +ΔCPD, where $V_{el}$ is electrostatic potential and ΔCPD is contact potential difference between the tip and the surface). Mapping the nulling potential $V_{dc}$ yields a surface potential map.

### 4.2.2. Domain Structure Reconstruction from SPM

The ability to image electrostatic as well as topographic properties of the surface allows surface domain structure to be reconstructed, as illustrated here for $BaTiO_3$ (100) surface. The tetragonal symmetry of $BaTiO_3$ results in characteristic surface corrugations at 90° *a-c* domain walls (provided that the crystal was polished in the cubic phase) as illustrated in Figure 1. The corrugation angle is $\theta = \pi/2 - 2\arctan(a/c)$, where $a$ and $c$ are the parameters of the tetragonal unit cell.[30] Thus, topographic imaging allows only *a-c* walls to be distinguished. The difference in electric properties of the surface due to the polarization charge, referred to as "potential" in Figure 1, allows *c*-domains of opposite polarity to be distinguished. An example of surface domain structure of $BaTiO_3$ is illustrated in Figure 2. The surface corrugations oriented in the *y*-direction indicate the presence of 90° *a-c* domain walls. The surface potential is uniform within *a* domains, whereas *c* domains exhibit both positive and negative regions, forming a number of 180° walls perpendicular to 90° domain boundaries (Figure 2a,c,e). This domain pattern can be ascribed to *c* domain wedges in the crystal with dominating *a* domain structure. The formation of 180° walls within the wedge minimizes the depolarization energy. If *c* domain regions are large (Figure 2b,d), irregular 180° walls separating $c^+$-$c^-$ domains exist.

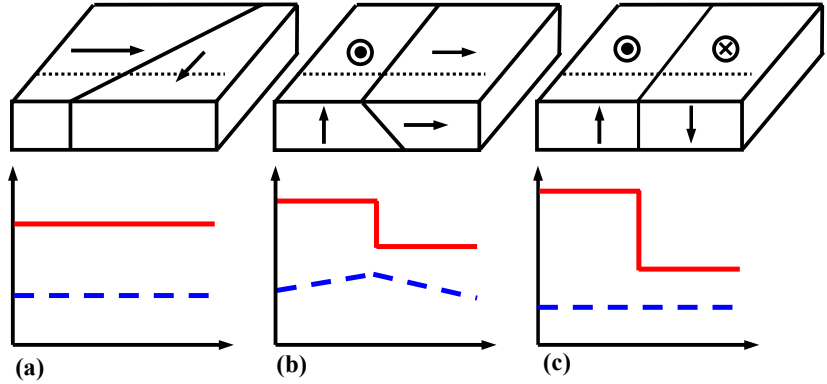

**Fig. 1.** Domain arrangements on (100) surface of tetragonal BaTiO$_3$. Arrows represent the orientations of polarization vectors. (a) 90° *a*1-*a*2 boundary, (b) 90° *c*$^+$-*a*1 boundary, (c) 180° *c*$^+$-*c*$^-$ boundary. Shown below is surface potential (solid) and surface topography (dashed) expected along the dotted line.

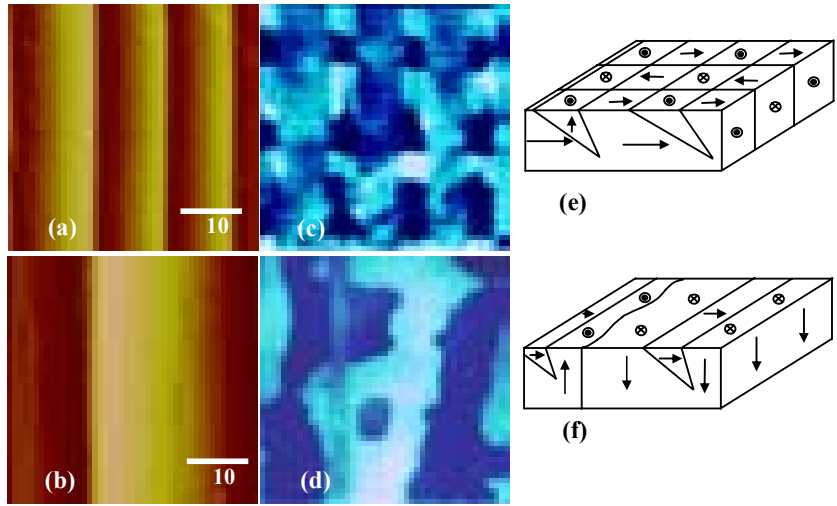

**Fig. 2.** Surface topography (a,b), surface potential (c,d) and schematics of domain structure (e,f) in *a*-domain region with *c*-domain wedges (a,c,e) and in *c*-domain region with *a*-domain wedges (b,d,f).

These walls are continuous through *a* domain regions, indicating the presence of *a* wedge domains in preferentially *c* domain material (Figure 2f).

SSPM image yields a potential difference between *c*$^+$ and *c*$^-$ domains as ~150 mV and between *a* and *c* domains as ~75 mV. The obvious question is related to the quantitative interpretation of domain potential contrast in terms of the polarization charge and the chemistry of the surface. Less obviously, even the qualitative relationship between domain potential and the polarization direction cannot be obtained from these data, i.e. positive or negative domains on potential image cannot be unambiguously associated with $c^+$ or $c^-$ polarization orientation and more detailed analysis of image formation mechanism in non-contact electrostatic SPM is required.

### 4.2.3. Image formation mechanism in electrostatic SPM

Ferroelectric domain contrast in non-contact SPMs is related to the polarization charge density difference between the domains having different out-of-plane component of polarization vector. It is recognized that in ambient polarization charge must be screened by surface electronic states or adsorption, reducing the depolarization energy of a multidomain ferroelectric.[25] To quantitatively address electrostatic properties of ferroelectric surfaces, the surface layer is represented with polarization charge $\sigma = \mathbf{P} \cdot \mathbf{n}$ and screening charge equivalent to surface charge density, $\sigma_s$, of the opposite polarity. For future discussion, it is convenient to introduce the degree of screening $\alpha = -\sigma_s/\sigma$. Partially or completely screened surfaces ($\alpha \leq 1$) are likely to be the equilibrium state of ferroelectric surfaces in air, whereas an overscreened surface ($\alpha > 1$) can occur during bias-induced domain switching.[31,32]

From the qualitative observations, both EFM and SSPM contrast is found to be uniform within the domains with rapid variation at the domain boundaries. The magnitude of potential and force gradient features are virtually domain-size independent. From the simple electrostatic arguments, this contrast can be attributed either to electrostatic field for an unscreened surface ($\alpha = 0$) or surface potential on a completely screened surface ($\alpha = 1$).[33]

To determine the origin of the EFM contrast, the distance and bias dependence of the average force gradient and the force gradient difference between domains was measured.[33] Based on the analysis of these data, the potential difference between $c^+$ and $c^-$ domains is determined as $\Delta V_{c-c} \approx 135 - 155$ mV. The average surface potential, $V_{av}$, is approximately equal to $(V_1+V_2)/2$, i.e. effective surface areas of $c^+$ and $c^-$ domain regions are equal, as expected from energy considerations. The potential difference between $a$ and $c^+$ domains was similarly found to be 85 mV, i.e. approximately equal to the expected value $\Delta V_{a-c} \approx \Delta V_{c-c}/2$. These results suggest that screening is symmetric, i.e. the degree of screening for $c^+$ and $c^-$ domains is the same.

The analysis of SSPM imaging mechanism is significantly complicated by a non-local cantilever contribution to the measured signal and by feedback effects as discussed elsewhere.[34] Direct measurement on BaTiO$_3$ (100) surface yielded potential variation between the antiparallel domains as $\Delta V_{c-c} \sim 130$ mV, which is remarkably close to the domain potential difference from complex and time con-suming analysis of EFM data (~150 mV), justifying the application of the former technique for ferroelectric characterization.

The potential difference between $c^+$ and $c^-$ domains $\Delta V_{c\text{-}c} \approx 150\text{mV}$ is equivalent to a 0.20 nm screening layer of a dielectric constant $\varepsilon = 80$ ($H_2O$). The alternative description involving screening by the charge carriers yields unphysical values for the depletion width (9.5 nm) for the undoped material with high dielectric constant. In addition, experimentally observed screening is symmetric. This is not the case if the screening is due to the free carriers in materials with a predominant electron or hole conduction, in which the width of accumulation layer for the polarization charge opposite to the majority carrier charge and width of depletion layer for the polarization charge similar to the majority carrier charge are vastly different. Therefore, preliminary analysis of SPM results indicate that the state of $BaTiO_3$ (100) surface under ambient conditions corresponds to almost complete screening of polarization bound charges.

### 4.2.4. Variable temperature SPM and charge dynamics

An additional insight into screening behavior on ferroelectric surfaces can be obtained from variable temperature measurements.

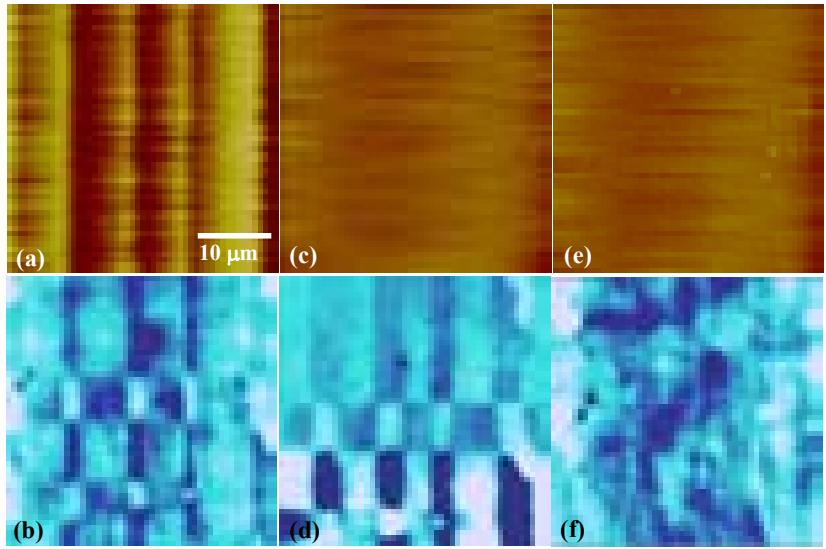

**Fig. 3.** Surface topography and potential distribution at $BaTiO_3$ (100) surface before ferroelectric phase transition at 125°C (a,b), 4 min after transition (c,d) and after 2.5 h annealing at 140°C (e,f). Scale is 0.1 V (b), 0.5 V (d) and 0.05 V (f).

Indeed, spontaneous polarization of ferroelectric is a function of temperature and disappears above Curie temperature. Thus, observation of temperature dependence of domain contrast can shed some light on surface behavior of ferroelectric material. In addition, such measurements allow in-situ observations of temperature dependent domain dynamics and ferroelectric phase transitions. An example of variable temperature imaging of domain structure in $BaTiO_3$ (100) surface is illustrated in Fig. 3. Above the Curie temperature, surface polarization disappears as indicated by the absence of surface corrugations. Unexpectedly, this is not the case for potential. The morphology of the potential features remains essentially the same (comp. Figure 3b,d), however, at the transition the potential amplitudes grow by almost 2 orders of magnitude. The relative sign of the domain potential features remains constant throughout the transition. As can be seen from Figure 3d (the image was acquired from bottom to top 4 min after the transition, total acquisition time – 11 min) the potential contrast decays with time. Surface potential distribution after remaining at 140°C for 2.5 h is shown in Figure 3f. The surface potential amplitude is now very small (~2-5 mV) and the potential distribution is almost random, though some resemblance to surface potential distribution below $T_c$ still exists. The magnitude of contrast of domain unrelated potential features remains unchanged.

Equally remarkable behavior is observed during the reverse process, i.e. cooling, as illustrated in Figures 4 and 5. Again, surface topography indicates the presence of 90° *a-c* domain walls. The SSPM image indicates the presence of domains of opposite polarity within *c* domains. On increasing the temperature the domain structure does not change significantly as illustrated in Fig. 4. Neither 90° nor 180° domain wall motion is observed. After a temperature decrease from 70°C to 50°C the domain contrast inverts (Fig. 5a,b), i.e. a positive *c* domain becomes negative. The potential difference between the domains decreases with time, passing through an isopotential point corresponding to zero domain potential contrast (Fig. 5c), and finally establishing an equilibrium value (Fig. 5d).

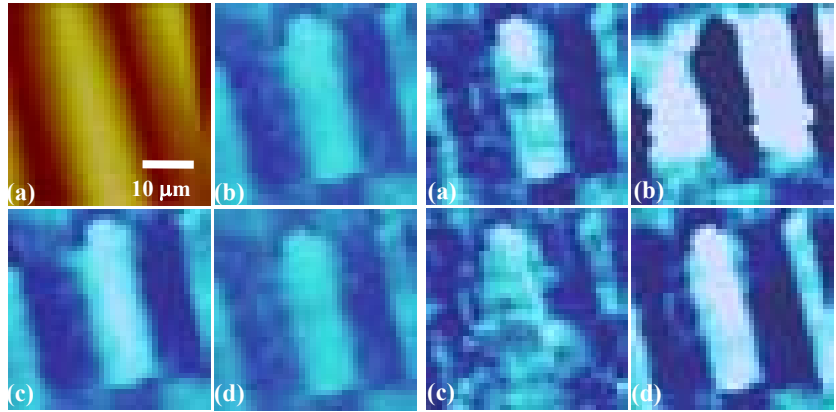

**Fig. 4.** Surface topography (a) and surface potential (b) of the ferroelectric domain structure on a $BaTiO_3$ (100) surface at *T* = 50°C. Surface potential after heating from 50°C to 70°C (c) and after annealing at 70°C for 50 min (d).

**Fig. 5.** Surface potential (a) of ferroelectric domain structure on $BaTiO_3$ (100) surface at T = 90°C. Surface potential during cooling from 90°C to 70°C (b), at 70°C (c) and after annealing at 70°C for 50 min (d).

The phenomena of potential retention above Curie temperature and temperature induced domain potential inversion allow a self-consistent description of screening on ferroelectric surfaces. In the case of complete screening, the surface potential has the sign of the screening charges and is reverse to that expected from polarization orientation, i.e. $c^+$ domains are negative and $c^-$ domains are positive on the SSPM image. Increasing the temperature results in a decrease of polarization charge, leaving some of the screening charge uncompensated and increasing the effective surface potential.

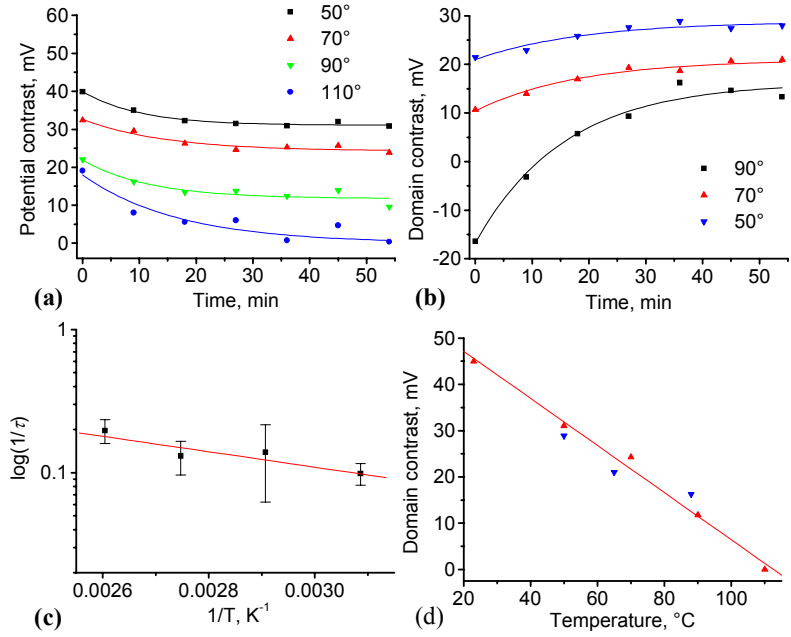

**Fig. 6.** Time dependence of domain potential contrast on heating (a) and cooling (b). Solid lines are fits by Eq. (4). (c) Time constant for relaxation process on heating in Arrhenius coordinates and (d) temperature dependence of equilibrium domain potential contrast on heating (▲) and cooling (▼).

On decreasing the temperature, spontaneous polarization increases and, for a short period of time, the sign of domain potential is determined by the polarization charge. Under isothermal conditions, polarization and screening charges equilibrate and the potential achieves an equilibrium value.

To analyze the temperature and time dependences of topographic structure and domain potential contrast the average corrugation angle and average domain potential difference between $c$ are $a$ domains were determined. The time dependence of domain potential contrast on heating and cooling is shown in Fig. 6a,b. To quantify the kinetics, the time dependence of domain potential contrast, $\Delta\varphi$, was approximated by an exponential function

$$\Delta\varphi = \Delta\varphi_0 + A\exp(-t/\tau), \qquad (4)$$

where $\tau$ is relaxation time and $A$ is a prefactor. Due to the finite heating and cooling rates, the domain potential contrast immediately after the temperature change can not be reliably established; therefore, Eq. (4) describes the late stages of potential relaxation. The temperature dependence of the potential redistribution time is shown in Fig. 6c. The redistribution time is almost temperature independent with an associated relaxation energy of ~4 kJ/mole. This low value of activation energy suggests that the kinetics of relaxation process is limited by the transport of charged species to the surface. The characteristic redistribution time is ~ 20 min and is close to the relaxation time for domain potential contrast above $T_c$ (30 min).[11]

The redistribution process both on heating and cooling results in the same equilibrium value of domain potential contrast, $\Delta\varphi_0$. The temperature dependence of domain potential contrast, shown in Fig. 6d, is almost linear, with the zero potential difference corresponding to temperature ~110°C well below the Curie temperature of $BaTiO_3$ ($T_c = 130$°C). For higher temperatures the degree of screening is smaller and the Coulombic contribution to the effective SSPM potential increases. Since polarization charge and screening charge contributions to the effective surface potential are of opposite sign, the decrease of the degree of screening results in the *decrease* of domain potential contrast. The thermodynamics of this process are expected to be strongly temperature dependent and to dominate over relatively weak variation of spontaneous polarization with temperature ($P = 0.26$ C/m$^2$ at 25°C and 0.20 C/m$^2$ at 100°C).

The equilibrium domain potential contrast can be related to the degree of screening of spontaneous polarization. As shown in Fig. 6d, the temperature dependence of equilibrium domain potential contrast in the temperature interval 30°C < $T$ < 100°C is linear and the domain potential contrast is the same on heating and cooling, i.e. equilibrium is achieved. This dependence can be represented by the linear function $\Delta V_{dc} = 0.059 - 5.3 \cdot 10^{-4} T$, where $T$ is temperature in Celsius degrees. The analysis of electrostatic tip-surface interaction[35] allows the temperature dependence of equilibrium degree of screening to be calculated as

$$1 - \alpha = 1.627 \cdot 10^{-5} + 1.23 \cdot 10^{-6} T \qquad (5)$$

From Eq.(5) and thermodynamic analysis of screening on ferroelectric surface, the enthalpy, $\Delta H_{ads}$, and entropy, $\Delta S_{ads}$, of adsorption are estimated as $\Delta H_{ads}$ = 164.6 kJ/mole, $\Delta S_{ads}$ = -126.6 J/mole K.[35]

The enthalpy and entropy of adsorption thus obtained are within expected values in spite of the approximations inherent in this approach. The Coulombic contribution to the effective potential can be estimated as < 10-20 % thus validating our previous conclusion that the surface is completely screened at room temperature. The nature of the screening charges can not be determined from these experiments; however, these results are also consistent with the well known fact that adsorbing on polar transition metal oxide surface in air are water and hydroxyl groups, -OH .[36,37,38] Dissociative adsorption of water as a dominant screening mechanism on $BaTiO_3$ surface in air was verified using temperature programmed desorption experiments on poled $BaTiO_3$ crystals.[39] Obviously, adsorbates can provide the charge required to screen the polarization bound charge, since corresponding polarization charge densities are of order of 0.25 $C/m^2$ corresponding to $2.6 \cdot 10^{-6}$ $mole/m^2$. For a typical metal oxide surface with characteristic unit cell size of ~ 4 Å this corresponds to the coverage of order of 0.25 mL.

## 4.3. Contact Imaging and Polarization Dynamics

### 4.3.1. Piezoresponse Force Microscopy and Lithography

Among the techniques for ferroelectric surface imaging listed in Table 1 the most widely used currently is Piezoresponse Force Microscopy, due to the ease of implementation, high resolution and relative insensitivity to topography and the state of the surface. It is not an exaggeration to say that PFM is rapidly becoming one of the primary imaging tools in the ferroelectric thin film research that routinely allows high resolution (~ 10 nm) domain imaging. In contrast to X-ray techniques, which are limited to averaged analysis of domain structure, PFM yields spatially resolved information on domain size, correlations, domain behavior near the inhomogeneities and grain boundaries. PFM can be used for imaging static domain structure in thin film, single crystals and polycrystalline materials, selective poling of specified regions on ferroelectric surface, studies of temporal and thermal evolution of domain structures, quantitative measurements of thermal phenomena and local hysteresis measurements. The information provided by PFM is summarized in Figure 7.

Despite the wide range of applications, contrast formation mechanism in PFM is still under debate.[40,41,42,43] The early treatments assumed that the measured electromechanical response is equal or proportional to the piezoelectric constant $d_{33}$ of the material, with some deviations due to the clamping by surrounding material. However, Luo et al.[12] have found that the temperature dependence of piezoresponse contrast is similar to that of the spontaneous polarization. This behavior was attributed to the dominance of electrostatic interactions due to the presence of polarization bound charge,[44] since the electromechanical response based on the piezoelectric coefficient, $d_{33}$, would diverge in the vicinity of the Curie temperature. Similar observations were reported by other groups.[14,45,46] The presence of electrostatic forces hypothesis is also supported by measurements on the nonpiezoelectric surfaces.[47] In contrast, the existence of a lateral PFM signal,[48,49,50] the absence of relaxation behavior in PFM contrast as opposed to SSPM contrast, as well as numerous observations using both EFM/SSPM and PFM[51,52] clearly point to significant electromechanical contribution to PFM contrast. Significant progress in the understanding of PFM was achieved recently.[45,53,54,55] In this section, contrast formation mechanism is analyzed and relative magnitudes of electrostatic vs. electromechanical contributions to PFM interaction for the model case of $c^+$, $c^-$ domains in tetragonal perovskite ferroelectrics are determined.

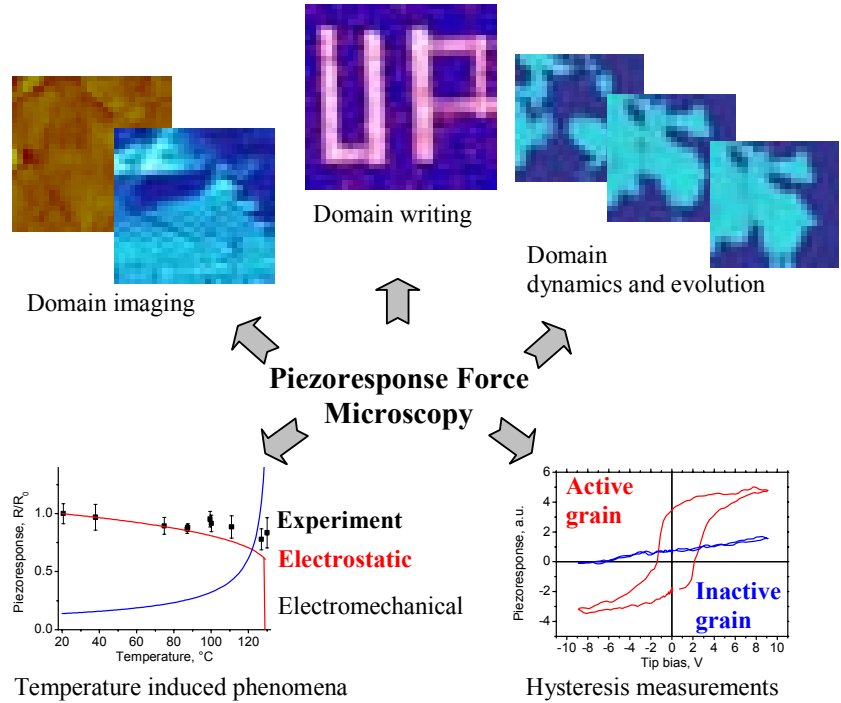

**Fig. 7.** Applications of Piezoresponse Force Microscopy

### 4.3.2. Principles of PFM

Piezoresponse Force Microscopy is based on the detection of bias-induced surface deformation. The tip is brought into contact with the surface and the piezoelectric response of the surface is detected as the first harmonic component of bias-induced tip deflection, $d = d_0 + A\cos(\omega t + \varphi)$. The phase of the electromechanical response of the surface, $\varphi$, yields information on the polarization direction below the tip. For $c^-$ domains (polarization vector pointing downward) the application of a positive tip bias results in the expansion of the sample and surface oscillations are in phase with the tip voltage, $\varphi = 0$. For $c^+$ domains $\varphi = 180°$. The piezoresponse amplitude, $PR = A/V_{ac}$, defines the local electromechanical activity of the surface. For a purely electromechanical response, piezoresponse amplitude is equal for $c^+$ and $c^-$ domains, while it is zero at the domain boundary. The width of the amplitude depression at the domain boundary is an experimental measure of the spatial resolution of the technique. The width of the phase profile cannot serve as the definition of the resolution; rather it represents the noise level and/or time constant of the lock-in amplifier.

One of the major complications in PFM is that both long range electrostatic forces and the electromechanical response of the surface contribute to the PFM signal so that the experimentally measured piezoresponse amplitude is $A = A_{el} + A_{piezo} + A_{nl}$, where $A_{el}$ is electrostatic contribution, $A_{piezo}$ is electromechanical contribution and $A_{nl}$ is non-local contribution due to capacitive cantilever-surface interactions.[44,48,56] Quantitative PFM imaging requires $A_{piezo}$ to be maximized to achieve predominantly electromechanical contrast. Alternatively, for $A_{el} \gg A_{piezo}$, electrostatic properties of the surface will be imaged. Often local polarization provides the dominant contribution to both electromechanical and electrostatic properties of the surface; qualitative imaging of domain structures is thus possible in both electromechanical and electrostatic cases. Cantilever size is usually significantly larger than domain size; therefore, a non-local cantilever contribution is usually present in the form of additive offset on PFM x-image. It can however lead to the erroneous interpretation of phase and amplitude images as shown in the Figure 8.

It is illustrative to estimate the effect of these interactions on PFM images using formalism developed by Hong *et. al.*[57] Assuming that $A_{el} = F_{loc}V_{ac}(V_{tip} - V_{loc})$, $A_{piezo} = d_{eff}V_{ac}$ and $A_{nl} = F_{nl}V_{ac}(V_{tip} - V_{av})$, where $V_{tip}$ is tip potential, $V_{loc}$ is local potential below the tip apex, $d_{eff}$ is effective electromechanical response of the surface, $V_{av}$ is average surface potential below the cantilever, $F_{loc}$ and $F_{nl}$ are proportionality coefficients determined by tip-surface and cantilever surface capacitance gradients, tip-surface contact stiffness and spring constant of the cantilever. In the simplest approximation (the field is uniform) $d_{eff}$ is equal to $d_{33}$; taking into account second order effects, $d_{eff} = d_{33} + Q(V_{tip} - V_{loc})$, where $Q$ is the corresponding electrostrictive coefficient.

In the purely electromechanical case, $F_{loc}$ and $F_{nl}$ are identically zero. In this case, the response amplitudes are equal in $c^+$ and $c^-$ domain regions, while phase changes by 180° between the domains. For domains with arbitrary orientation, the absolute value of the amplitude signal provides a measure of the piezoelectric activity of the domain; in-plane domains or non-ferroelectric regions are seen as the regions with zero response amplitude. PFM spectroscopy yields hysteretic behave-ior from which materials properties such as piezoelectric and electrostriction coefficients can be obtained.

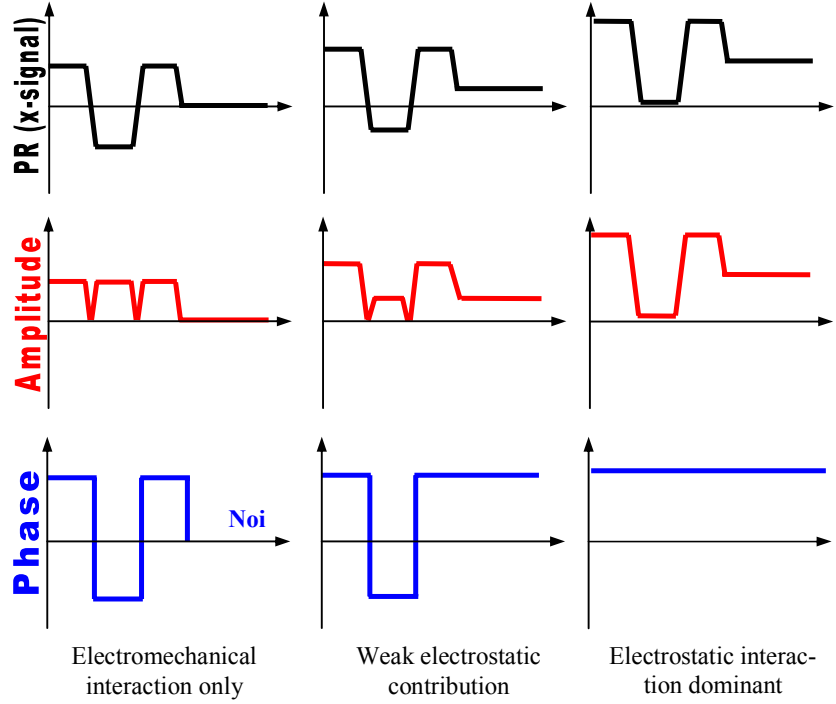

**Fig. 8.** Schematic of signals in PFM of $c^+/c^-/c^+/a$ domain structure. Shown are *x*-signal, amplitude and phase signal for purely electromechanical case, with weak electrostatic contribution and strong electrostatic contribution.

In the more realistic case, both electrostatic and electromechanical interactions contribute and the PFM *x*-signal over $c^+$ and $c^-$ domains can be written as:

$$PR_+ = d_{33} + Q(V_{tip} - V_1) + F_{loc}(V_{tip} - V_1) + F_{nl}(V_{tip} - V_{av}), \qquad (6a)$$

$$PR_- = -d_{33} + Q(V_{tip} - V_2) + F_{loc}(V_{tip} - V_2) + F_{nl}(V_{tip} - V_{av}). \qquad (6b)$$

For a small non-local electrostatic contribution, $d_{33} \gg F_{nl}(V_{tip} - V_{av})$, the phase still changes by 180° between the domains (Figure 8); however, the response amplitudes are no longer equal in $c^+$ and $c^-$ domain regions. Non-ferroelectric regions are seen as the region with finite response amplitude. Similar behavior is expected for non-zero local electrostatic contribution, in which case the piezoresponse is a sum of electromechanical and electrostatic contributions and depends linearly on tip bias, $PR_\pm = d_{33} \pm F_{loc}(V_{tip} - V_{loc})$. The immediate implication of Eqs.(6 a,b) is that second order electrostriction coefficients can be determined by PFM if and only if imaging is purely electromechanical; otherwise, non-local electrostatic response is measured as discussed in Section 4.3.6. For a large non-local contribution, $F_{nl}(V_{tip} - V_{av}) \gg d_{33}$, the phase is determined by electrostatic force only and does not change between the domains. The amplitude signal is strongly asymmetric and maximal response corresponds to either $c^+$ or $c^-$ domain, while minimal signal is observed to the domain with opposite polarity. In-plane domains or non-ferroelectric regions are seen as regions with intermediate contrast on amplitude image. Obviously, even qualitative analysis of PFM image in this case is difficult: for example, a sample comprised of $a$-$c^+$ and $c^-$-$c^+$ domains can not be distinguished based on a vertical PFM image only.

As follows from Eqs.(6 a,b), quantitative spectroscopic piezoresponse measurements require that the electrostatic and non-local components of the response be minimized. The first step in developing a systematic approach is the reliable calculation of the magnitudes of electrostatic, electromechanical and non-local responses as a function of tip radius of curvature, indentation force and cantilever spring constant.

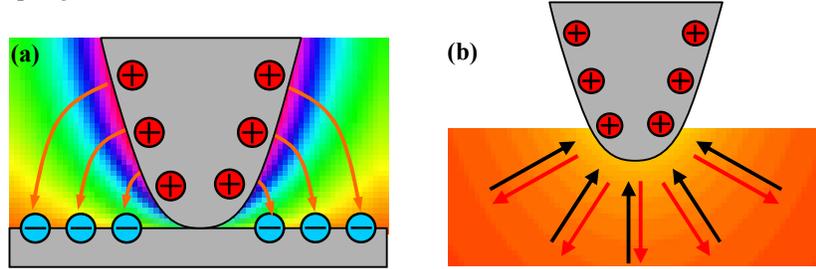

**Fig. 9.** Electrostatic (a) and electromechanical (b) contrast in PFM

One of the difficulties in a comparison of the relative magnitudes of electromechanical and electrostatic responses is the difference in the contrast transfer mechanism. In the electromechanical case, the surface displacement due to inverse piezoelectric effect is determined as a function of applied voltage (Figure 9). Tip deflection in this case is equal to surface displacement since the contact stiffness of the tip–surface junction is usually much larger than cantilever spring constant. In the electrostatic case, the force containing both local and non-local components is defined. In the subsequent sections, the influence of imaging conditions on PFM contrast is analyzed. The magnitudes of $A_{el}$ and $A_{piezo}$ are determined as a function of indentation force and tip radius of curvature. The case when the measured response amplitude owes to electrostatic tip-surface interaction only, $A = A_{el}$, is further referred to as electrostatic regime. The case when the electromechanical contribution dominates, $A = A_{piezo}$, is referred to as electromechanical regime. The non-local contribution, $A_{nl}$, is governed by the cantilever spring constant rather than indentation force and is considered separately in Section 4.6.3.

### 4.3.3. Electrostatic Contrast in PFM

In the electrostatic regime of piezoresponse force microscopy the capacitive and Coulombic tip-surface interactions result in an attractive force between the tip and the surface, that causes surface indentation below the tip.

*Potential Distribution in the Tip-surface Junction*

The potential distribution in the tip-surface junction in non-contact imaging is often analyzed in the metallization limit for the surface.[58] In this limit, the tip-surface capacitance $C_d(z,\kappa)$, where $z$ is the tip-surface separation and $\kappa$ is the dielectric constant for the sample is approximated as $C_d(z,\kappa) \approx C_c(z)$, where $C_c(z)$ is the tip-surface capacitance for a conductive tip and conductive surface. This approximation breaks down for small tip-surface separations when the effect of field penetration in the dielectric sample is non-negligible. For ferroelectric surfaces, the effective dielectric constant is high, $\kappa \approx 100\text{-}1000$, favoring the metallization limit. However, in contact tip-surface separation $z \approx 0$ leads to a divergence in the capacitance, $C_c(z)$, and the corresponding force. To avoid this difficulty and, more importantly, take into account the anisotropy of the ferroelectric medium, we calculate the tip-surface force using the image charge method for spherical tip geometry. This approach is applicable when the tip-surface separation is small, $z \ll R$, where $R$ is radius of curvature of the tip.

**Table 2.** Images for conductive, dielectric and anisotropic dielectric planes

|    | Conductive | Isotropic dielectric | Anisotropic dielectric |
|----|------------|----------------------|-------------------------|
| $Q'$ | $-Q$ | $-\dfrac{\kappa-1}{\kappa+1}Q$ | $-\dfrac{\sqrt{\kappa_z\kappa_x}-1}{\sqrt{\kappa_z\kappa_x}+1}Q$ |
| $d'$ | $-d$ | $-d$ | $-d$ |
| $Q''$ | $0$ | $\dfrac{2\kappa}{\kappa+1}Q$ | $\dfrac{2\sqrt{\kappa_z\kappa_x}}{\sqrt{\kappa_z\kappa_x}+1}Q$ |
| $d''$ |  | $d$ | $d\sqrt{\kappa_z/\kappa_x}$ |

The potential in air produced by charge $Q$ at a distance $d$ above a conductive or dielectric plane located at $z = 0$ can be represented as a superposition of potentials produced by the original charge and the corresponding image charge $Q'$ located at position $z = d'$ below the plane. The potential in a dielectric material is equal to that produced by a different image charge $Q''$ located at $z = d''$.[59,60,61] Values of $Q'$, $Q''$, $d'$ and $d''$ for metal and isotropic or anisotropic dielectric materials are summarized in Table 2. Note that the potential in air above an anisotropic dielectric mate-rial is similar to the isotropic case with an effective dielectric constant $\kappa_{eff} = \sqrt{\kappa_x \kappa_z}$, where $\kappa_x$, $\kappa_z$ are the principal values of the dielectric constant tensor. This simple image solution can be used only for the dielectric for which one of the principal axes of dielectric constant tensor coincides with surface normal and in-plane dielectric constants are equal, $\kappa_x = \kappa_y$. For systems without in-plane isotropy, field distribution is no longer rotationally invariant and does not allow representation with single image charge; more complex solutions using distributed image charges are required.[62]

To address tip-surface interactions and taking the effect of the dielectric medium into account, the image charge distribution in the tip can be represented by charges $Q_i$ located at distances $r_i$ from the center of the sphere such that:

$$Q_{i+1} = \frac{\kappa-1}{\kappa+1} \frac{R}{2(R+d)-r_i} Q_i \quad (7a)$$

$$r_{i+1} = \frac{R^2}{2(R+d)-r_i} \quad (7b)$$

where $R$ is tip radius, $d$ is tip-surface separation, $Q_0 = 4\pi\varepsilon_0 RV$, $r_0 = 0$ and $V$ is tip bias. Tip-surface capacitance is

$$C_d(d,\kappa) = \frac{1}{V}\sum_{i=0}^{\infty} Q_i, \quad (8)$$

from which the force can be found. The rotationally invariant potential distribution in air can be found from Eqs.(7 a,b). One of the important parameters for the description of tip-surface junction is potential on the surface directly below the tip, which defines the potential attenuation in the tip-surface gap. Specifically, for sphere plane model potential on the surface directly below the tip is

$$V(0,0) = \frac{1}{4\pi\varepsilon_0} \frac{2}{\kappa+1} \sum_{i=0}^{\infty} \frac{Q_i}{R+d-r_i}. \quad (9)$$

In the conductive surface limit, $\kappa = \infty$ and Eq.(8) is simplified to[63]

$$C_c = 4\pi\varepsilon_0 R \sinh\beta_0 \sum_{n=1}^{\infty} (\sinh n\beta_0)^{-1}, \quad (10)$$

where $\beta_0 = \operatorname{arccosh}((R+d)/R)$. Surface potential in this case is $V(0,0) \equiv 0$. For the conductive tip-dielectric surface

$$C_d = 4\pi\varepsilon_0 R \sinh\beta_0 \sum_{n=1}^{\infty} \left(\frac{\kappa-1}{\kappa+1}\right)^{n-1} (\sinh n\beta_0)^{-1}. \quad (11)$$

While in the limit of small tip-surface separation $C_c$ diverges logarithmically, $C_d$ converges to the universal "dielectric" limit[64]

$$C_d(\kappa)_{z=0} = 4\pi\varepsilon_0 R \frac{\kappa-1}{\kappa+1} \ln\left(\frac{\kappa+1}{2}\right). \qquad (12)$$

The distance dependence of tip-surface capacitance and surface potential directly below the tip are shown in Figure 10a,b.

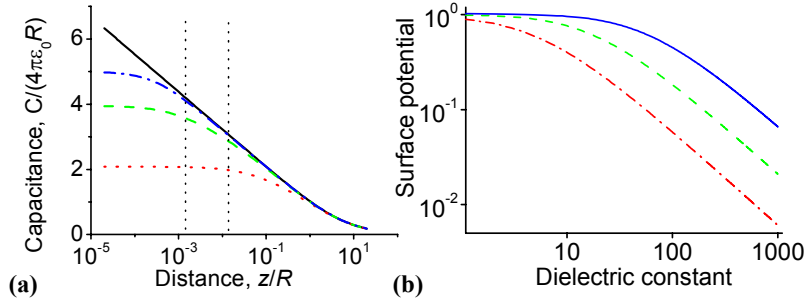

**Fig. 10.** (a) Tip-dielectric surface capacitance for $\kappa = 10$ (dot), $\kappa = 100$ (dash) and $\kappa = 1000$ (dash-dot), compared to the metallic limit (solid line). Vertical lines delineate the region of characteristic tip-surface separations (0.1-1nm) in contact mode for tip radius $R = 50$nm. (b) Surface potential below the tip for tip-surface separations $z = 0.1\ R$ (dash-dot), $z = 0.01\ R$ (dash) and $z = 0.001\ R$ (solid line) as a function of the dielectric constant of the surface.

For relatively large tip-surface separations $C_d(z,\kappa) \approx C_c(z)$, which is the usual assumption in non-contact SPM imaging. The most prominent feature of this solution is that, while for low-$\kappa$ dielectric materials the tip-surface capacitance achieves the dielectric limit in contact and hence surface potential is equal to the tip potential, this is not the case for high-$\kappa$ materials. Tip-surface capacitance, capacitive force and electric field can be significantly smaller than in the dielectric limit. The surface potential below the tip is smaller than the tip potential and is inversely proportional to dielectric constant (Figure 10b). This is equivalent to the presence of an apparent dielectric gap between the tip and the surface that attenuates the potential. Even though under typical experimental conditions contact area between the tip and the surface is non-zero and the sphere-plane model is not rigorous, this analysis can still be used. For the clean ferroelectric surface, the gap is due to intrinsic phenomena such as finite Thomas-Fermi length (for metallic tips) or Debye length (for semiconductor tips) of the tip material (~0.5 A) and/or non-uniform polarization distribution in surface layer of ferroelectric. For a small contact area and a large dielectric constant the potential drop in such an intrinsic gap can be significant. In many practical cases, surface damage, contamination or loss of volatile components during fabrication result in an extrinsic non-ferroelectric dead layer. Eq.(9) and Figure 10 illustrate the implications of a dielectric gap on tip-induced surface potential since the minimal tip-surface separation, $d$, is limited by dead layer width.

*Tip-surface Forces and Indentation Depth*

The electrostatic contribution, $A_{el}$, to piezoresponse amplitude is calculated assuming that the total force acting on the tip is comprised of the elastic contribution due to the cantilever, $F_0 = kd$, and a capacitive force, $F_{el} = C'(V_{tip} - V_{surf})^2$, where $k$ is the cantilever spring constant, $d$ the setpoint deflection, $C'$ is the tip-surface capacitance gradient and $V$ is the potential. From Eqs.(10,11), the magnitudes of capacitive and Coulombic forces between the cantilever-tip assembly and the surface can be estimated. The capacitive force is:

$$2F_{cap} = C'_{loc}(V_{tip} - V_{loc})^2 + C'_{nl}(V_{tip} - V_{av})^2, \qquad (13)$$

where $V_{tip}$ is the tip potential, $V_{loc}$ is the domain-related local potential directly below the tip, $V_{av}$ is the surface potential averaged over the distance comparable to the cantilever length, $C'_{loc}$ is the local part of tip-surface capacitance gradient and $C'_{nl}$ is the non-local part due to the cantilever. Typically, the cantilever length is significantly larger than the characteristic size of ferroelectric domains; therefore, the non-local part results in a constant background on the image that does not preclude qualitative domain imaging. The non-local capacitance gradient can be estimated using plane-plane geometry as $C'_{nl} = \varepsilon_0 S(z+L)^{-2}$, where $S$ is the effective cantilever area and $L$ is the tip length. For a typical tip with $L \approx 10$ μm and $S \approx 2 \cdot 10^3$ μm$^2$, the non-local contribution is $C'_{nl} \approx 1.8 \cdot 10^{-10}$ F/m and is independent of tip radius. The force for a tip-surface potential difference of 1 V is $F_{nl} \approx 0.9 \cdot 10^{-10}$ N. Here, it is assumed that the force acting on the cantilever results in surface indentation only and the cantilever geometry does not change, i.e. tip displacement is measured. Practically, the optical detection scheme used in most modern AFMs implies that the cantilever bending angle is measured. Bias induced buckling oscillations of the cantilever, which are not associated with vertical tip motion, result in the effective displacement as analyzed in Section 4.6.3.

The local capacitive contribution due to the tip apex is $F_{loc} = 1.4 \cdot 10^{-8}$ N for $z$ = 0.1 nm, $R$ = 50 nm, i.e. two orders of magnitude larger. However, $C'_{loc}$ scales linearly with tip radius and, therefore, for the sharp tips capable of high-resolution non-local contributions to the signal increase. Similar behavior is found for non-contact SPMs.[65] The Coulombic tip-surface interaction due to polarization charge can be estimated using the expression for the electric field above a partially screened ferroelectric surface, $E^u = (1-\alpha)P\varepsilon_0^{-1}(1+\sqrt{\kappa_x \kappa_z})^{-1}$, where α is the degree of screening and $P$ is spontaneous polarization ($P$ = 0.26 C/m$^2$ for BaTiO$_3$). For unscreened surfaces $\alpha = 0$ so this Coulombic contribution in the limit $F_{coul}$<<$F_{cap}$ is $F_{coul} = C_{loc}(V_{tip} - V_{loc})E^u$ and for the same tip parameters as above $F_{coul} = 2.2 \cdot 10^{-9}$ N. However, polarization charge is almost completely screened in air ($1-\alpha \ll 10^{-3}$), and under these conditions the Coulombic contribution can be excluded from the electrostatic tip-surface interaction for isothermal experiments.

Capacitive force results in an indentation of the surface. In the Hertzian approximation the relationship between the indentation depth, $h$, tip radius of curvature, $R$, and load, $P$, is[66]

$$h = \left(\frac{3P}{4E^*}\right)^{2/3} R^{-1/3}, \qquad (14)$$

where $E^*$ is the effective Young's modulus of the tip-surface system defined as

$$\frac{1}{E^*} = \frac{1-\nu_1^2}{E_1} + \frac{1-\nu_2^2}{E_2}. \qquad (15)$$

$E_1$, $E_2$ and $\nu_1$, $\nu_2$ are Young's moduli and Poisson ratios of tip and surface materials. For ferroelectric perovskites the Young's modulus is of the order of $E^* \approx$ 100GPa. The contact radius, $a$, is related to the indentation depth as $a = \sqrt{hR}$. Hertzian contact does not account for adhesion and capillary forces in a tip-surface junction and a number of more complex models for nanoindentation processes are known.[67]

Under typical PFM operating conditions the total force acting on the tip is $F = F_0 + F_{el}$, where $F_0 = k\,d_0$ is elastic force exerted by the cantilever of spring constant $k$ at setpoint deflection $d_0$ and $F_{el}$ is the electrostatic force. Since the electrostatic force is modulated, $V_{tip} = V_{dc} + V_{ac}\cos(\omega t)$, the first harmonic of tip deflection is

$$h_{1\omega} = \frac{\chi}{2\pi\omega}\int \left(F_0 + C'_{loc}(V_{dc} + V_{ac}\cos(\omega t) - V_{loc})^2\right)^{\frac{2}{3}} \cos(\omega t)\,dt, \qquad (16)$$

where $\chi = (3/4E^*)^{2/3} R^{-1/3}$. In the limit when the indentation force is much larger than electrostatic force, $F_{el} \ll F_0$, the effective spring constant of the tip-surface junction is $k_{eff} = \partial P/\partial h$ and the first harmonic of cantilever response is $h_{1\omega} = F_{1\omega}/k_{eff}$. For a Hertzian indentation the response is:

$$h_{1\omega} = \frac{2}{3}\left(\frac{3}{4E^*}\right)^{2/3} R^{-1/3} F_0^{-1/3} F_{1\omega}. \qquad (17)$$

This equation can be also obtained directly from an expansion of the integrand in Eq.(16). For typical PFM imaging conditions the setpoint deflection is ~ 100 nm and the spring constant of the cantilever $k$ varies from ~0.01 to ~100 N/m. Consequently, imaging can be done under a range of loads spanning at least 4 orders of magnitude from 1 nN to 10 μN. For $F_0 = 100$ nN, $E^* = 10^{11}$ Pa and potential difference between the domains $\Delta V = 150$ mV, PFM contrast between the domains of opposite polarities is $\Delta h_{1\omega} = 6.02 \cdot 10^{-12}$ m/V. It should be noted that the potential difference between ferroelectric domains in ambient is determined by the properties of the adsorbate layer that screens spontaneous polarization.[35] Under UHV conditions where intrinsic screening by charge carriers[25] dominates the potential difference would be larger and can achieve the limiting value of $\Delta V = 3$ V comparable to band gap. In this case, the electrostatic PFM contrast between the domains of opposite polarities can be as large as $\Delta h_{1\omega} = 1.2 \cdot 10^{-10}$ m/V.

### 4.3.4. Electromechanical Contrast

The analysis of electrostatic interactions above is applicable to any dielectric surface; however, for ferroelectric and, more generally, piezoelectric materials an additional bias-induced effect is a linear electromechanical response of the surface. A rigorous mathematical description of the problem is extremely complex and involves the solution of coupled electromechanical mixed boundary value problem

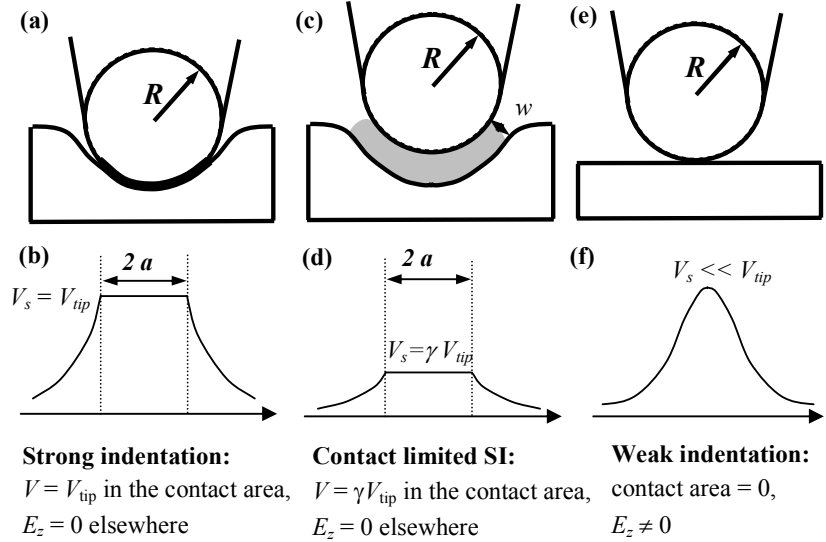

**Fig. 11.** Limiting cases for the electromechanical interactions in the PFM. Tip-surface junction (a,c) and surface potential (b,d) in the strong indentation limit with and without the apparent gap effect and tip-surface junction (e) and surface potential (f) in the weak indentation limit.

for the anisotropic medium. Fortunately, the geometry of the tip-surface junction in PFM is remarkably similar to the well-studied piezoelectric indentation problem.[68,69,70,71,72,73] In the classical limit, the coupled electromechanical problem is solved for mixed value boundary conditions; $V_s = V_{tip}$ in the contact area and the normal component of the electric field $E_z = 0$ elsewhere, as illustrated in Figure

11. It was shown in Section 4.3.3 that surface potential can be significantly attenuated due to the dielectric tip-surface gap. The gap can originate both from the intrinsic properties of tip-surface junction and due to the presence of "dead layers" on the ferroelectric surface.[74] To account for this effect, we introduce semiclassical contact limited strong indentation, in which $V_s = \gamma V_{tip}$ in the contact area, where $\gamma$ is the attenuation factor. SPM experiments can also be performed under the conditions when the contact area is negligibly small. Piezoelectric deformation occurs even when the tip is not in contact due to the tip-induced non-uniform electric field. In this case, the zero field approximation outside of the contact area is invalid; instead, the contact area itself can be neglected and the surface deformation can be ascribed solely to field effects. Therefore, we distinguish two limits for the PFM electromechanical regime:

1. Strong (classical) indentation: $V = V_{tip}$ in the contact area, $E_z = 0$ elsewhere
2. Weak (field induced) indentation: contact area is negligible, $E_z \neq 0$

In practice, both mechanisms might operate and the dominant contribution depends on imaging parameters. Interestingly, by this definition, response calculated in the weak indentation (WI) model allows the error associated with $E_z = 0$ assumption in the strong indentation (SI) model to be estimated.

*Strong Indentation Limit*

A complete description of the strong indentation limit is given by Suresh and Giannakopoulos,[75,70] who extended Hertzian contact mechanics to piezoelectric materials. Assuming the continuous strain at the contact edge, $\sigma_{zz}(a,0) = 0$, the relationship between the load, $P$, indentor potential, $V$, and indentation depths, $h$, is

$$h = \frac{a^2}{R} + \frac{2\beta}{3\alpha}V, \qquad (18a)$$

$$P = \alpha \frac{a^3}{R} \qquad (18b)$$

where $\alpha$ and $\beta$ are materials dependent constants and $a$ is contact radius. Eq. (18a) yields piezoresponse as $d_{si} = 2\beta/3\alpha$. An alternative form of Eqs.(18a,b) was obtained by Giannakopoulos and Suresh for different continuity condition at the contact edge, $du_z(a,0)/dr = const$ (smooth deformation transition). In this case, the response amplitude is non-linear function of indentor bias as analyzed elsewhere.[76] The response amplitude in the strong indentation limit is comparable to the corresponding $d_{33}$ value (Table 3). The difference between $d_{si}$ and $d_{33}$ illustrates the clamping effect due to the difference of field distribution below the tip and in the uniformly biased material.

The applicability of the strong indentation regime to PFM contrast is limited. A high dielectric constant leads to a significant potential drop between the tip and the surface, $V_s \ll V_{tip}$; therefore, for an infinitely stiff tip and surface the basic assumption of the strong indentation limit, $V_s = V_{tip}$ in the contact area, is not fulfilled. Even for finite contact the potential on the surface below the tip is lower than the tip potential and differs from that assumed in the strong indentation limit. It is useful to consider the effect of contact radius on this assertion. A simple approximation for the surface potential below the tip is $V_s = \gamma V_{tip}$ in the contact area, where γ is the attenuation factor (Figure 11c,d). Such behavior is referred to as contact limited strong indentation (CSI). Using a spherical approximation for contact region, the attenuation factor is estimated as $\gamma = (1 + w\kappa_{eff}/a\kappa_d)^{-1}$, where $w$ is the thickness of the "apparent" dielectric gap ($w > 0.1$ nm), $\kappa_d$ is the dielectric constant in the gap ($\kappa_d = 1$-$100$), $a$ is the contact radius and $\kappa_{eff}$ is the effective dielectric constant of the ferroelectric material. For planar geometry (i.e. $R >> a >> w$), $\kappa_{eff}$ is close to $\kappa_z$ for a ferroelectric material. For the spherical case, $\kappa_{eff}$ is close to $\sqrt{\kappa_x \kappa_z}$, imposing an upper and lower limit on $\kappa_{eff}$. For a metallic tip, the gap effect is expected to be minimal; for a perfect conductor the lower limit on the effective tip-surface separation is set by the Thomas-Fermi length of the metal. For doped silicon tips $w$ will be comparable to the depletion width of the tip material. Even for thin dielectric layers (0.1 nm-1 nm) the effective surface potential can be attenuated by as much as a factor of 100 due to a large difference between the dielectric constants of gap layer and ferroelectric. For imperfect contact, the magnitude of the piezoresponse in the strong indentation limit can become comparable to that of the electrostatic mechanism. The deviation of the tip shape from spherical (e.g. flattening due to wear, etc) reduces the electrostatic response due to a higher contact stiffness and increases electromechanical response. The resolution in the strong indentation limit is limited by indentation radius $a$.

### *Weak Indentation Limit*

Weak indentation defines the other limiting case in the PFM experiment when the indentation load and contact area are small. In this limit, the contribution of the contact area to the total electromechanical response of the surface can be neglected (Figure 11e,f). The potential distribution in the tip-surface junction is calculated ignoring the electromechanical coupling (rigid electrostatic limit) as shown in Section 4.3.3 since the dielectric constant of material is high and field penetration into ferroelectric is minimal. The electromechanical response of the surface is calculated using the Green's function for point force/charge obtained by Karapetian et al.[72]

$$h(r) = f\frac{A}{r} + q\frac{L(s_{ij}, e_{ij}, \varepsilon_{xx}, \varepsilon_{zz})}{r}, \tag{19}$$

where $h$ is vertical displacement, $r$ is the radial coordinate, $f$ is the point force, $q$ is the point charge, $A$ and $L$ are materials dependent constants and $r$ is the distance from the indentation point. For distributed charge, the surface deflection is:

$$h(\mathbf{r}) = L\int \frac{\sigma(\mathbf{r}_0)}{|\mathbf{r}-\mathbf{r}_0|}dS, \quad \text{where} \quad \sigma(\mathbf{r}_0) = \varepsilon_0 E_z(\mathbf{r}_0) \tag{20}$$

The materials properties affect the PFM contrast through the coefficient *L*, while the geometric properties are described by the (materials independent) integral. This treatment assumes that the field penetration into the material is small.

For spherical tip geometry, the electromechanical surface response in the weak indentation limit can be evaluated using the image charge method. The surface charge density induced by point charge *Q* at distance *l* from a conductive or high-$\kappa$ dielectric surface is $\sigma_0 = \frac{Q}{4\pi} \frac{2d}{(l^2 + r^2)^{3/2}}$. From Eq.(20), charge induced piezoelectric deformation of the surface is $h = QL/l$. Using the image charge series developed in Section 4.3.3, total tip-induced surface deformation is

$$h = L \sum_{i=0}^{\infty} \frac{Q_i}{R + d - r_i} = LG(R,d). \tag{21}$$

Note that this expression is remarkably similar to that of the tip-induced surface potential [Eq.(9)]. Thus, piezoresponse in the weak indentation limit can be related to tip-induced surface potential, $V_s$, as $h = 2\pi\varepsilon_0 L(\kappa + 1)V_s$. Specifically, the surface deformation is linear in surface potential, $h = d_{wi}V_s$, where the effective piezoresponse constant, $d_{wi}$, in the weak indentation limit is $d_{wi} = 2\pi\varepsilon_0 L(\sqrt{\kappa_x \kappa_z} + 1)$ The characteristic numerical values for *L* and $d_{wi}$ are summarized in Table 3.

**Table 3.** Piezoresponse coefficients for different materials

| Composition | $d_{33}$, m/V | $d_{si}$, m/V | *L*, m$^2$/C | $D_{wi}$, m/V |
|---|---|---|---|---|
| BaTiO$_3$ | 1.91·10$^{-10}$ | 1.70·10$^{-10}$ | 1.54·10$^{-3}$ | 1.10·10$^{-10}$ |
| PZT4 | 2.91·10$^{-10}$ | 2.48·10$^{-10}$ | 2.41·10$^{-3}$ | 1.71·10$^{-10}$ |
| PZT5a | 3.73·10$^{-10}$ | 3.02·10$^{-10}$ | 2.66·10$^{-3}$ | 2.05·10$^{-10}$ |

For *R* = 50 nm, *d* = 0.1 nm and a typical value of $L \approx 2.5 \cdot 10^{-3}$ m$^2$/C the characteristic piezoresponse amplitude in the weak indentation limit is $h \approx 6.54 \cdot 10^{-12}$ m/V. The distance and tip radius dependence of the response is $h \sim (R/d)^{0.5}$, in agreement with a previously used point charge approximation.[77] The effective piezoelectric constant $d_{wi}$ for weak indentation limit is remarkably similar to $d_{si}$ for the strong indentation limit as shown in Table 3. The difference between the limits arises from the disparate ways the dielectric gap is taken into account (Figure 11). The weak indentation limit accounts for the effect of the gap directly in the functional form of coefficient *L*, which incorporates the dielectric properties of the surface. In the strong indentation limit, the effective dielectric gap must be introduced through the attenuation factor, $\gamma$. The resolution in the weak indentation limit is determined by the tip radius of curvature and effective tip-surface separation and is proportional to $\sqrt{Rh}$.

### *Effect of Materials Properties on Piezoresponse*

The solutions discussed above present a mathematically rigorous description of PFM contrast mechanism. These solutions clearly illustrate that complete analysis of the electromechanical response of the surface in terms of materials properties is difficult. Even in the ideal case of known geometry, both strong and weak indentation limits lead to complex expressions that include 9 out of 10 electroelastic constants for a transversally isotropic medium. For the systems with lower symmetry (e.g. ferroelectric grain with random orientation) the analytical treatment of the problem is even more complex. Therefore, the understanding of PFM contrast can be greatly facilitated if a simplified relationship between piezoresponse and material properties can be established. As illustrated in Table 3, in many cases the effective response calculated from rigorous models is comparable to the $d_{33}$ of the material. Given the difference between the geometries of the problems ($d_{33}$ defines the electromechanical response in *z*-direction to the uniform field applied in the *z*-direction, effective piezoresponse defines the electromechanical response in the *z*-direction to the highly non-uniform field below the tip), these results are quite surprising. In order to rationalize this observation, here we investigate the contribution of various electromechanical constants to piezoresponse in the strong and weak indentation limits.

Electroelastic properties of the solid can be described either in terms of elastic compliances $s_{ij}$ [m$^2$/N], piezoelectric constants $d_{ij}$ [C/N or m/V] and dielectric permittivities $\varepsilon_{ij}$ [F/m], or by elastic stiffness constant $c_{ij}$ [N/m$^2$], piezoelectric constants $e_{ij}$ [C/m$^2$ or Vm/N] and dielectric permittivities $\varepsilon_{ij}$ [F/m].[78] These sets of constants are related by the tensorial relations $d_{nj} = e_{ni}s_{ij}$, $e_{nj} = d_{ni}c_{ij}$, $s_{ij} = c_{ij}^{-1}$ and $c_{ij} = s_{ij}^{-1}$. In order to clarify the relative contributions of different electroelastic constants to PFM, responses in both the strong and weak indentation limits are calculated for a variety of ferroelectric materials[78,79,80] A sensitivity function of piezoresponse is defined as the functional derivative of piezoresponse with respect to material parameter, $S(f_{ij}) = \delta PR(f_{ij})/\delta f_{ij}$. The sensitivity is calculated numerically as

$$S(f_{ij}) = \frac{PR(f_{ij} = 1.1 f_{ij}^0) - PR(f_{ij} = 0.9 f_{ij}^0)}{0.2 PR(f_{ij} = f_{ij}^0)}, \tag{22}$$

where $f_{ij}$ is a selected electroelastic constant and $f_{ij}^0$ is a reference value for that constant. A positive value of $S(f_{ij})$ implies that a higher constant favors piezoresponse, while a negative value of $S(f_{ij})$ suggests that the response decreases with this constant. $S(f_{ij}) \approx 0$ indicates that the response is independent of the property.

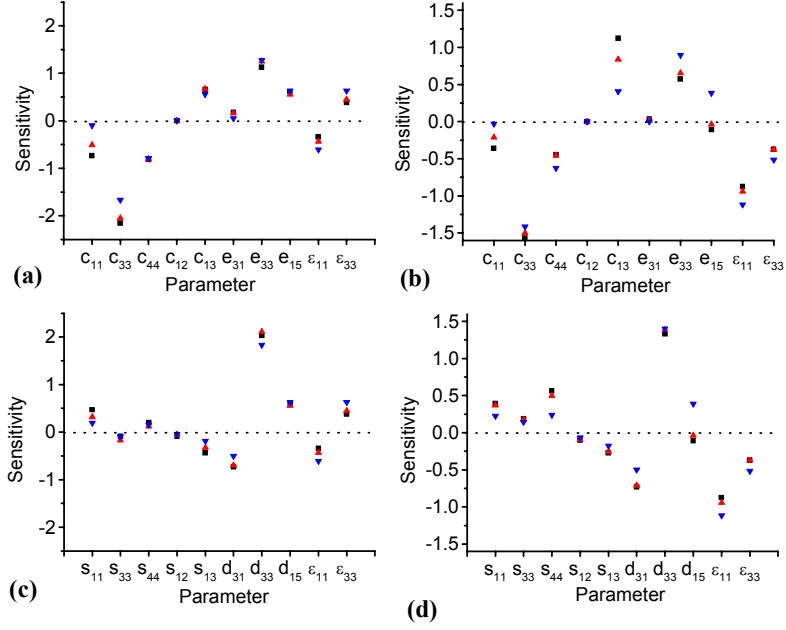

**Fig. 12.** Sensitivity in the strong (a) and weak (b) indentation limits in the in the ($c_{ij}$, $e_{ij}$, $\varepsilon_{ij}$) representation. Shown are sensitivities for BaTiO$_3$ (■), PZT4 (▲) and PZT5a (▼). Sensitivity in the strong (c) and weak (d) indentation limits in the in the ($s_{ij}$, $d_{ij}$, $\varepsilon_{ij}$) representation. Shown are sensitivities for BaTiO$_3$ (■), PZT4 (▲) and PZT5a (▼).

Figure 12a,b shows the sensitivity of piezoresponse in the strong indentation limit to electroelastic constants in the ($c_{ij}$, $e_{ij}$, $\varepsilon_{ij}$) representation. Note that response decreases for larger elastic stiffnesses, and the largest contribution originates from $c_{33}$. As expected, response increases for larger piezoelectric constants $e_{ij}$. The contributions of various electroelastic constants in this representation to piezoresponse are comparable. Sensitivity of piezoresponse for several ferroelectric materials in the ($s_{ij}$, $d_{ij}$, $\varepsilon_{ij}$) representation is shown in Figure 12 c,d. Piezoresponse in the strong indentation limit is clearly dominated by the $d_{33}$ of the material, while other electroelastic constants provide minor contributions (Figure 12c). This observation implicitly justifies the often used assumption of measured piezoresponse being equal to $d_{33}$. In the weak indentation limit, both $d_{33}$ and $\varepsilon_{11}$ strongly influence the response, significant contributions being provided by $d_{31}$ and $\varepsilon_{33}$ as well (Figure 12 d). The response increases with $d_{33}$ and decreases with $\varepsilon_{11}$ as expected. The response in both limits does not depend on elastic stiffness $c_{12}$ (Figure 12 c,d).

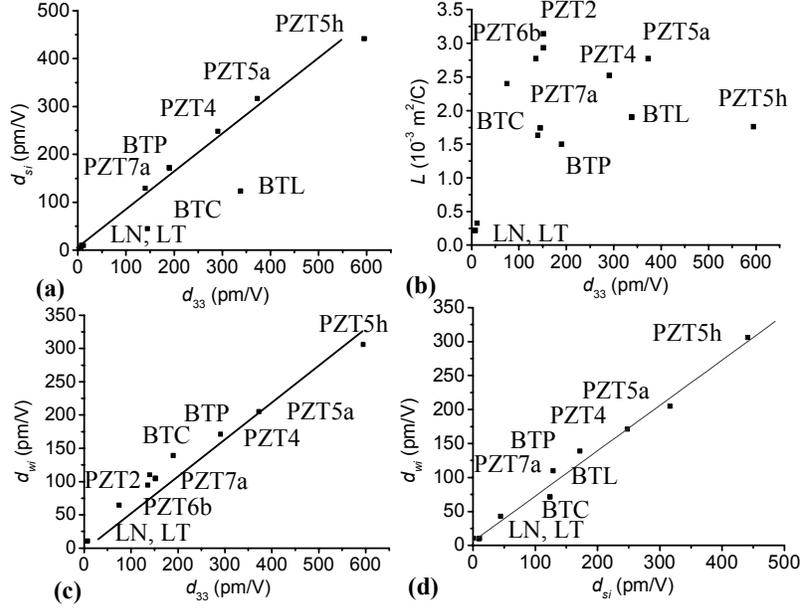

**Fig. 13.** Correlation between piezoresponse and $d_{33}$ in the strong (a) and weak (b) indentation limits for some polycrystal and single crystal materials. Correlation between piezoelectric constant $d_{wi}$ and $d_{33}$ (c) and correlation between $d_{wi}$ and $d_{si}$ (d). PZT denotes different types of commercial lead zirconate-titanate ceramics, LN and LT are $LiNbO_3$ and $LiTiO_3$, BTC is 95%$BaTiO_3$/5%$CaTiO_3$ (ceramic B), BTP and BTL are $BaTiO_3$ polycrystals.

The goal is to determine under what conditions a correlation exists between the measured piezoresponse and $d_{33}$ of the material. Earliest treatments of piezoresponse image contrast explicitly assumed that the response is proportional or equal to $d_{33}$. To test this assertion, the calculated piezoresponse coefficient is compared to piezoelectric constant for a series of ferroelectric materials. An almost linear correlation exists between the response in the strong indentation limit and $d_{33}$, $d_{si} \sim 0.75\, d_{33}$ (Figure 13a). In contrast, no such correlation is observed between $L$ and $d_{33}$ for the weak indentation limit (Figure 13b). The physical origin of this behavior is that $L$ defines the response of the surface to charge and therefore depends on ratios of the type $d_{ij}/\varepsilon_{ij}$. According to the Ginzburg-Devonshire theory these ratios are proportional to the corresponding second order electrostriction coefficients, $d_{ij}/\varepsilon_{ij} \sim Q_{ij}P$. Therefore, the effect of the electromechanical coupling coefficient and dielectric constants counteract each other. On the other hand, the effective piezoelectric constant in the weak indentation limit, $d_{wi}$, exhibits a good correlation with $d_{33}$, $d_{wi} \sim 0.5\, d_{33}$ (Figure 13c), since the dielectric constant effect is already accounted for. The effective piezoelectric response constants in the weak and strong indentation limits exhibit almost perfect linear dependence, $d_{wi} = 0.66$ $d_{si}$ (Figure 13d). This similarity is due to the fact that in the first approximation the piezoresponse is defined by surface potential directly below the tip; the minor differences in the proportionality coefficient between piezoresponse and $d_{33}$ are due to the differences in surface potential profile, as shown in Figure 11. The difference in the mechanisms is that in the strong indentation limit the *potential* in the contact area is assumed to be known and equal to the tip potential; the exact tip geometry is therefore not essential as long as contact is good. In the weak indentation limit the surface deflection is defined by tip-induced *charge distribution* on the surface, which strongly depends on tip geometry. If "true" PFM is the ability to quantify piezoelectric coefficient directly from the measurements, it can be achieved directly only in the strong indentation region. In the weak indentation regime, the electromechanical properties of the surface can be obtained indirectly provided that tip-surface geometry is known.

### 4.3.5. PFM Contrast Mechanism Maps

In PFM, the contrast mechanism will depend on details of the experimental conditions. Depending on tip radius and indentation force, both linear and non-linear electrostatic interactions and strong and weak indentation regimes can occur. In order to relate PFM imaging mechanisms to experimental conditions, Contrast Mechanism Maps were constructed as shown in Figure 14.[76,81] To delineate the regions with dominant interactions, surface deformation in the electrostatic case was estimated using the distance dependence of tip-surface capacitance as $F_{1\omega} = 2.7 \cdot 10^{-8} (R/50)(0.1/d)(V_{tip} - V_s)V_{ac}$ N, where both $R$ and $d$ are in nanometers. The surface deformation, $h_{1\omega}^{el}$, was calculated from Eq.(17). The boundaries of the non-local regions are established by a comparison of tip apex-surface capacitance and cantilever-surface capacitance.[82] Surface deformation in the electromechanical regime was calculated including the "apparent dielectric gap" effect as $h_{1\omega}^{em} = d_{eff} / (1 + w\kappa_{eff} / a\kappa_d)$, where contact radius, $a$, is given by the Hertzian model and $\kappa_{eff}/\kappa_d = 30$. The boundary between strong and weak indentation regimes is given by attenuation factor of 0.3. It should be noted that the estimate of the attenuation factor is the major source of uncertainty in this treatment; the Green function based description as described in Section 6.4.3 is expected to overcome this difficulty. The boundary between the electromechanical and the electrostatic regions is given by the condition $h_{1\omega}^{em} = h_{1\omega}^{el}$. For small indentation forces, a non-linear dynamic behavior of the cantilever is expected and the corresponding condition is $F_{el} = F_0$. For very large indentation forces, the load in the contact area can be sufficient to induce plastic deformation of the surface or the tip. The onset of this behavior is expected when $F_0 / \pi a^2 = E^*$. High pressures in the contact area can significantly affect the ferroelectric properties of material and induce local polarization switching, etc.[83,84,85] at a strain $P/d_{33} \sim 3\cdot 10^9$ N/m² for a typical ferroelectric material.

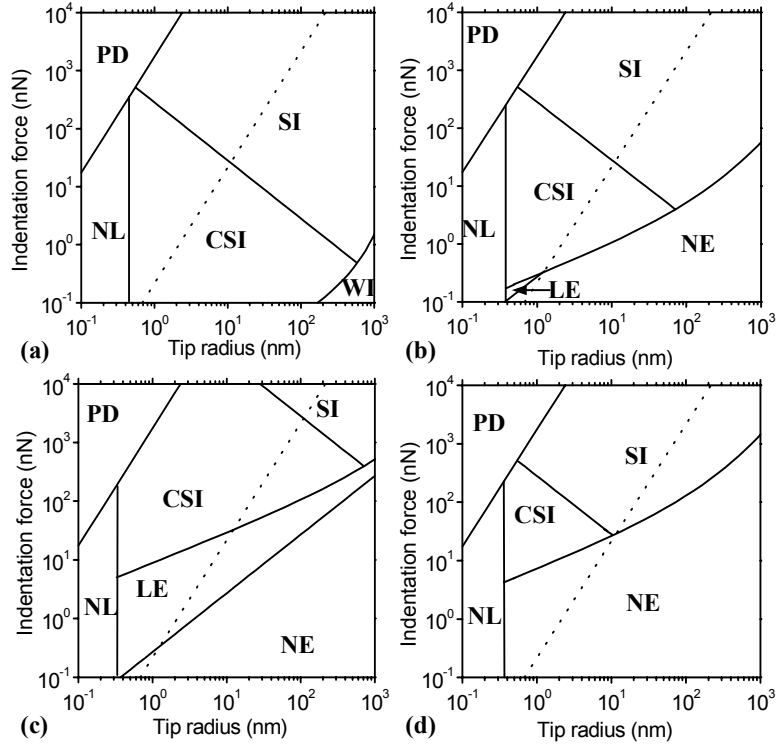

**Fig. 14.** Contrast Mechanism Maps of piezoresponse force microscopy. SI is strong indentation regime, CSI - contact limited strong indentation, WI - weak indentation regime, LE - linear electrostatic regime, NE - nonlinear electrostatic regime, NL - non-local interactions, PD - plastic deformation. The dotted line delineates the region where stress-induced switching is possible. (a) $w = 0.1$ nm, $\Delta V = V_{tip}-V_s = 0$ V, (b) $w = 0.1$ nm, $\Delta V = 1$ V, (c) $w = 1$ nm, $\Delta V = 1$ V, (d) $w = 0.1$ nm, $\Delta V = 5$ V.

The Contrast Mechanism Map in Figure 14a corresponds to imaging under good tip-surface contact ($w = 0.1$ nm) and zero tip-surface potential difference. The crossover from contact limited strong indentation to strong indentation limit depends on the choice of the attenuation factor. Pure weak indention behavior is observable only for large tip radii and small indentation forces. Typically, the ferroelectric domains are associated with surface potential variations and tip potential is not equal to surface potential. The Contrast Mechanism Map in Figure 14b corresponds to imaging under good tip-surface contact ($w = 0.1$ nm) and mod-erate tip-surface potential difference ($V_{tip}$-$V_{loc}$ = 1 V). Less perfect contact that results from oxidized tips or poorly conductive coating, as well as the presence of contaminants will expand the weak indentation and linear electrostatic regions, primarily at the expense of the strong indentation region (comp. Figure 14b,c). Increasing the tip-surface potential difference increases the electrostatic contribution (Figure 14d). Consequently, the non-linear electrostatic region expands and can even eliminate the linear electrostatic region. However, above a certain tip-surface potential difference or driving voltage the linear approximation Eq.(17) is no longer valid and Eq.(16) must be used. The effect of high driving voltages and tip-surface potential difference is an increase of indentation force $F = F_0 + C'_{loc}(V_{tip} - V_{loc})^2$, expanding the electromechanical region. Piezoelectric coefficient can be quantified directly from the measurements only in the strong indentation region. As shown in Figure 13, $d_{si}$ correlates linearly with $d_{33}$ in the strong indentation regime. In the weak indentation regime and contact limited strong indentation regime, the properties of the surface can still be obtained indirectly provided that the tip geometry is known. Finally, in the electrostatic regime, the PFM image is dominated by long-range electrostatic interactions and piezoelectric properties of material are inaccessible. In certain cases surface charge distribution is directly correlated with ferroelectric domain structure; therefore, qualitative information on domain topology can still be obtained. These results allow multiple controversies in the interpretation of PFM contrast to be reconciled by elucidating experimental conditions under which electrostatic vs. electromechani-cal mechanisms dominate. Acquisition of quantitative information requires blunt tips and intermediate indentation forces to avoid pressure-induced polarization switching, i.e. operation regimes to the right of the dotted line in Figure 14. The use of top metallic electrode as proposed by Christman[86] or liquid electrode as proposed by Ganpule[54] is the limiting case of this consideration.

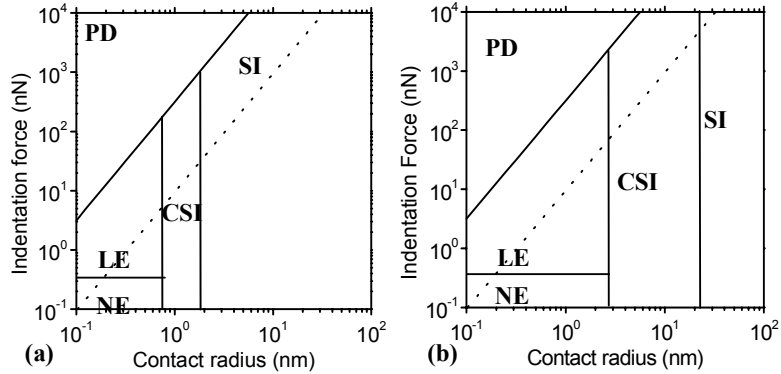

**Fig. 15.** Contrast Mechanism Maps of piezoresponse force microscopy as a function of contact radius and indentation force. SI corresponds to strong indentation regime, CSI - contact limited strong indentation, WI - weak indentation regime, LE - linear electrostatic regime, NE - nonlinear electrostatic regime, PD - plastic deformation. Dotted line delineates the region where stress-induced switching is possible. The maps are constructed for good tip-surface contact ($w$ = 0.1 nm) and bad contact ($w$ = 1 nm).

The Contrast Mechanism Maps in Figure 14 are semiquantitative for a spherical tip; however, gradual tip wear during the imaging is inevitable and can be easily detected using appropriate calibration standards. The influence of tip flattening on PFM contrast mechanisms is shown in Figure 15a,b. The response was calculated as a function of contact radius for fixed electrostatic force corresponding to $R = 100$ nm. In contrast to the spherical case, the contact stiffness for a flat indentor does not depend on the indentation force; hence, the crossover from the electrostatic to electromechanical regime occurs at some critical contact radius. Since the sphere/plane model is less accurate for this case, the degree of approximation associated with it results in the more qualitative nature of the contrast map. It should be noted, however, that electrostatic force can be measured directly[87] and used for the construction of the map for individual tip.

### 4.3.6. Non-local Effects: Cantilever Contribution to PFM

The non-local contribution to PFM, $A_{nl}$, arises due to the buckling oscillations of the cantilever[88] induced by capacitive cantilever-surface interactions as illustrated in Figure 16.[57] Typically, the cantilever length is significantly larger than the size of ferroelectric domains; therefore, the non-local interaction results in a constant background that does not preclude quantitative domain imaging but heavily contributes to local hysteresis measurements. In order to calculate the effective displacements due to the buckling oscillations, the simple harmonic-oscillator type models are no longer applicable; instead, realistic cantilever geometry must be taken into account.

Cantilever oscillations can be described by the beam equation[64]

$$\frac{d^4 u}{dx^4} + \frac{\rho A}{EI}\frac{d^2 u}{dt^2} = \frac{q(x,t)}{EI}, \tag{23}$$

where $E$ is the Young's modulus of cantilever material, $I$ is the moment of inertia of the crossection, $\rho$ is density, $A$ is cross-section area, and $q(x,t)$ is the distributed force acting on the cantilever. For a rectangular cantilever $A = wt$ and $I = wt^3/12$, where $w$ is cantilever width and $t$ is thickness. For a uniform periodic force Eq.(23) is solved by introducing $u(x,t) = u_0(x)e^{i\omega t}$, $q(x,t) = q_0 e^{i\omega t}$, where $u_0$ is displacement amplitude, $q_0$ is load per unit length, $t$ is time and $\omega$ is modulation frequency. After substitution Eq.(23) is:

$$\frac{d^4 u_0}{dx^4} = k^4 u_0 + \tilde{q}, \tag{24}$$

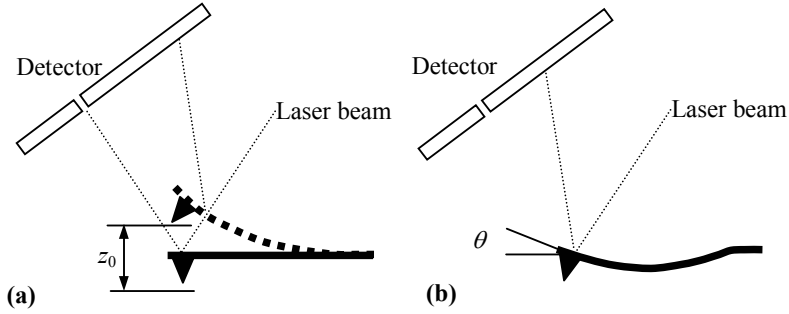

**Fig. 16.** Schematic diagram of non-local cantilever effect in PFM. Displacement of laser beam induced by cantilever deflection (a) is equivalent to that due to cantilever buckling induced by uniformly distributed load (b).

where $k^4 = \omega^2 \rho A/EI$, $\tilde{q} = q_0/EI$. The boundary conditions for Eq.(24) are $u_0(0) = 0$ and $u_0'(0) = 0$ on the clamped end and $u_0(L) = 0$, $u_0''(L) = 0$ on the supported end, where $L$ is cantilever length. Eq.(24) is solved in the usual fashion. Of interest is the deflection angle $\theta$ at $x = L$, which is related to the local slope as $\theta = \mathrm{atan}(u_0'(L)) \approx u_0'(L)$. The frequency response of effective deflection for the buckling oscillations of supported cantilever is

$$\theta = \frac{\tilde{q}(\cos(kL) - \cosh(kL) + \sin(kL)\sinh(kL))}{k^3(\cosh(kL)\sin(kL) - \cos(kL)\sinh(kL))}. \quad (25)$$

Resonant frequencies for cantilever oscillations are found as a solution of

$$\cosh(kL)\sin(kL) = \cos(kL)\sinh(kL). \quad (26)$$

The lowest order solutions of Eq.(26) are $\beta_n = kL = 3.927, 7.067, 10.21$. Corresponding eigenfrequencies are $\omega_n^2 = EI\beta_n^4/\rho SL^4 = Et^2\beta_n^4/12\rho L^4$. In comparison, for the free oscillating cantilever the frequency response is given by

$$\theta = \frac{\tilde{q}(\sinh(kL) - \sin(kL))}{k^3(1 + \cosh(kL)\cos(kL))}, \quad (27)$$

and the resonance occurs for

$$\cos(kL)\cosh(kL) + 1 = 0, \quad (28)$$

Several lowest order solutions of Eq.(28) are $\alpha_n = kL = 1.875, 4.694, 7.855$. Therefore, the first cantilever buckling resonance in contact mode occurs at ~ 4.4 times higher frequency than the resonance of the free cantilever.

In the low frequency limit Eq.(25) is simplified by $\theta = -L^3 q_0/48EI = -L^3 q_0/4wt^3 E$ and the effective oscillation amplitude detected by an optical detector is $2\theta L/3$.[29] For a freely oscillating cantilever in the low frequency limit, the response is larger by the factor of 8.

The capacitive cantilever-surface force is $F_{cap} = \varepsilon_0 S(V_{tip} - V_{surf})^2/2L^2$, where S is the cantilever area $S = Lw$. Therefore, the first harmonic of the load is $q_0 = \varepsilon_0 S V_{ac} \Delta V/2LH^2$, where $H$ is tip height equal to cantilever-surface separation and $\Delta V = V_{dc} - V_{surf}$. The non-local contribution to PFM signal is conveniently rewritten in terms of the spring constant of the free oscillating cntilver, $k_{eff} = Ewt^3/4L^3$ as $A_{nl} = -Lw\varepsilon_0 V_{ac} \Delta V/48 k_{eff} H^2$. Therefore, the piezoresponse signal in local hysteresis loop measurements comprising both electromechanical and non-local electrostatic parts is

$$PR = d_{eff} + \frac{Lw\varepsilon_0 (V_{dc} - V_{surf})}{48 k_{eff} H^2}, \qquad (29)$$

where the first term is the sum of electromechanical and local electrostatic contribution and the second term is due to the cantilever buckling oscillations.

As discussed above, PFM imaging and quantitative piezoresponse spectroscopy requires that the electromechanical interaction be much stronger than that of the non-local electrostatic interaction. From Eq.(29) the non-local contribution is inversely proportional to the cantilever spring constant, while the electromechanical contribution is spring constant independent. This condition can be written as $k_{eff} \gg k^* = Lw\varepsilon_0 \Delta V/48 d_{eff} H^2$, where $k^*$ is the critical cantilever spring constant corresponding to the equality of non-local cantilever-surface and electromechanical tip-surface interactions. Taking an estimate $d_{eff}$ = 50 pm/V, $\Delta V$ = 5 V, $L$ = 225 μm, $w$ = 30 μm, $H$ = 15 μm, the condition on the spring constant is $k_{eff}$ > 0.55 N/m. This condition can be easily modified for cantilevers with different geometric properties and can be rewritten as a condition for tip-surface potential difference. Note that while for $\Delta V$ = 0 non-local interactions are formally absent, this condition is hardly achieved experimentally unless a top-electrode set-up is used. Even though for the cantilever with high spring constants ($k_{eff}$ = 50 N/m) the electrostatic contribution is ~1% of electromechanical, it will hinder the determination of electrostriction coefficient from the saturated part of hysteresis loop.

The non-local contribution to PFM is illustrated in Figure 17, which compares local hysteresis loops obtained using cantilevers with large ($k$ = 5 N/m) and small spring constants ($k$ = 0.1 N/m). Both cantilevers allow successful PFM imaging since relative domain contrast in not influenced by the non-local contribution. However, only the stiff cantilever yields a well-defined local hysteresis loop. The soft cantilever exhibits a response linear in voltage due to the dominance of capacitive cantilever-surface force and cantilever buckling. Still, the contribution of electrostatic interactions is non-negligible for the first cantilever, as well, and can be detected on non-ferroelectric grains (Grain II). Note that the stiffness of the cantilever cannot be increased indefinitely: for a very stiff cantilever and a large indentation force materials properties (e.g. pressure induced polarization reversal or mechanical surface stability) limit the imaging as illustrated in the Figure14. For small indentation force the spring constant of tip-surface junction will become smaller than cantilever spring constant; hence tip deflection will be much smaller than the surface deflection.

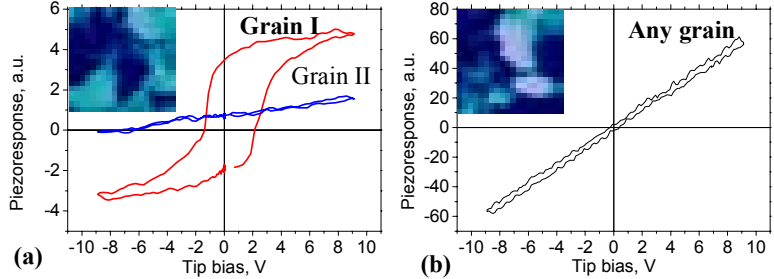

**Fig. 17.** Piezoresponse hysteresis loops for stiff (a) and soft (b) cantilevers. Upper insets show 1 μm scans of the surface verifying that imaging is possible in both cases.

### 4.3.7. Phase Transition and Polarization Dynamics by VT PFM

One of the origins of the existing ambiguity between electrostatic and electromechanical response mechanisms in PFM is the weak temperature dependence of experimentally measured piezoresponse. Here we apply the analytical solutions developed above to rationalize the temperature dependence of the piezoresponse of $BaTiO_3$. To distinguish the atomic polarization from surface potential, the phase transition was studied by PFM. The surface topography and piezoresponse at various temperatures are displayed in Figure 18. Surface corrugations indicate the presence of 90° *a-c* domain boundaries. The piezoresponse image reveals 180° domain walls separating regions of opposite polarity within *c*-domains. On heating from room temperature to 125°C the overall domain structure remains constant, however, small nuclei of domains of inverse polarity (Figure 18b,d) grow with temperature. On transition to the paraelectric state, both the surface corrugations and the piezoresponse contrast almost disappear. It should be noted that extremely weak inverted piezoresponse contrast could be observed after the transition (Figure 18f). This phenomenon is ascribed to the weak electrostatic interaction between the screening charges and SPM probe. On reverse transition, domain-related corrugations form very quickly (Figure 19c). Piezoresponse variation during the transition is complex, but clear piezoresponse contrast develops after the transition (Figure 19d) and after equilibration below $T_c$, a new well-defined *a-c* domain structure is established (Figure 19e,f).

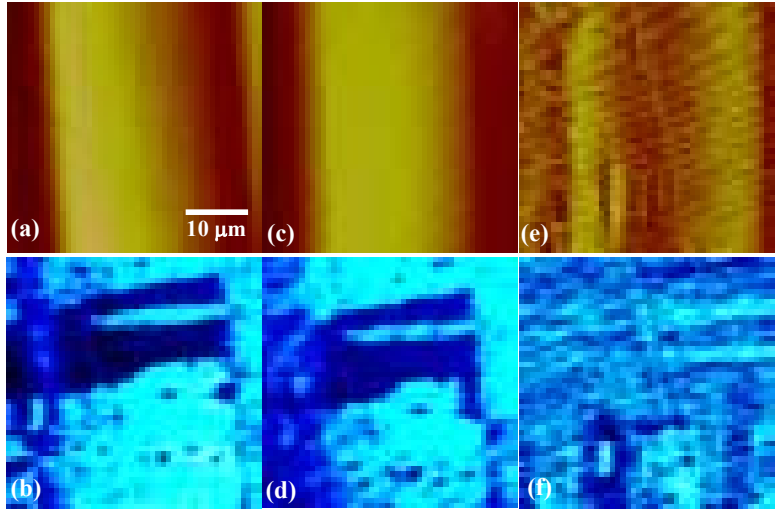

**Fig. 18.** Surface topography (top) and piezoresponse (bottom) of BaTiO$_3$ (100) surface before ferroelectric phase transition at 20°C (a,b), at 125°C (c,d) and 4 min after transition at 140°C (e,f). Scale is 30 nm (a,c,e).

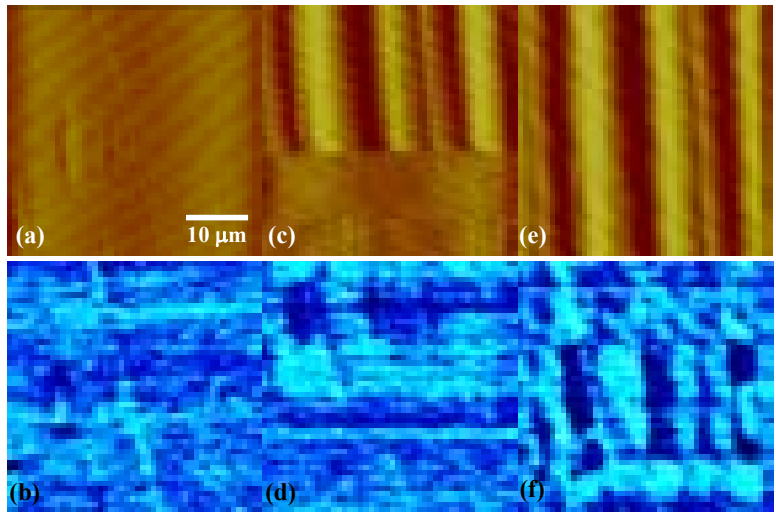

**Fig. 19.** Surface topography (top) and piezoresponse (bottom) of BaTiO$_3$ (100) surface above Curie temperature at 140°C (a,b), during the reverse ferroelectric phase transition at 130°C (c,d) and after 30 min annealing at 120°C (e,f). Scale is 30 nm (a,c,e).

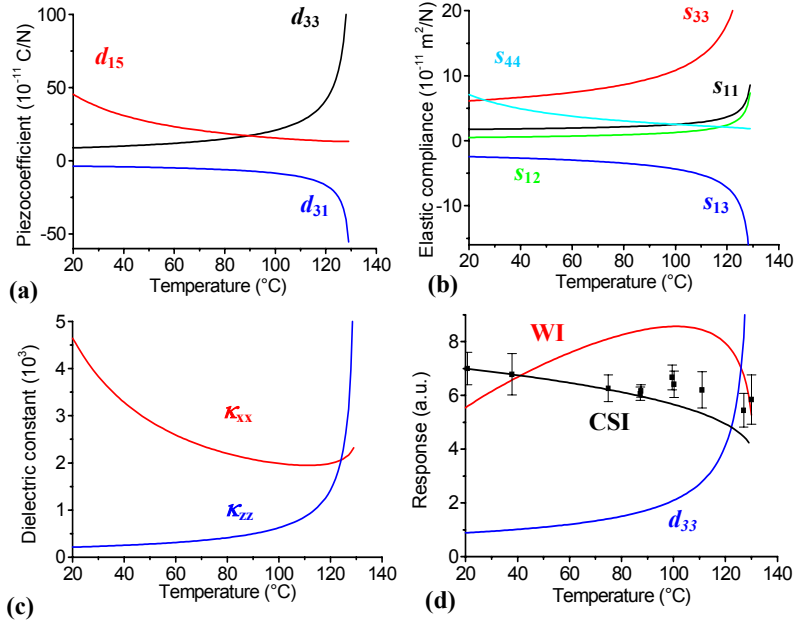

**Fig. 20.** Temperature dependence of elastic constants (a), piezoelectric constants (b) and dielectric constants (c) for $BaTiO_3$ calculated from Ginzburg-Devonshire theory and temperature dependence of piezoresponse coefficient in the WI and CSI limits (d).

Experimentally measured temperature dependence of piezoresponse contrast is illustrated in Figure 20. This temperature dependence is reminiscent of that of polarization and, indeed, capacitive interaction between the conductive tip and polarization charge has been used to describe the piezoresponse imaging mechanism. However, SSPM imaging suggests that polarization bound charge is completely screened on the surface.[14,33,35] Potential dynamics are extremely complex and exhibit relaxation behavior, which is not observed for the PFM signal. In order to explain the observed phenomena, we calculate the temperature dependence of PFM signal using the models developed above.

The temperature dependence of PFM contrast is calculated according to Karapetian *et. al.*[72] for weak indentation limit. The temperature dependence of the electroelastic constants for $BaTiO_3$ was calculated by Ginzburg-Devonshire theory[89,90] and the temperature dependence for $L(T)$ is compared to experimental measurements in Figure 20. In contrast to the strong indentation limit, no divergence occurs in the temperature dependence of the weak indentation and contact limited strong indentation limits, consistent with experimental behavior. The physical origin of this behavior is that not only the piezoelectric constant, but also the dielec-tric constant increases with temperature. The simplified model for this behavior can be developed assuming that $PR \sim V_{eff} d_{33}$, where $V_{eff}$ is the potential on the ferroelectric surface. For high attenuation of the tip-potential in the dielectric gap, $V_{eff} \approx a\varepsilon_d / (d\varepsilon_{33}) V_{tip}$. Taking into account that $d_{33} = 2\varepsilon_0 \varepsilon_{33} Q_{11} P$, the temperature dependence of piezoresponse can be calculated as

$$R \sim 2\varepsilon_0 \varepsilon_d Q_{11} P \frac{a}{d} V_{tip} \sim P, \qquad (30)$$

i.e. the temperature dependence of piezoresponse is that of spontaneous polarization. It is noteworthy that the predicted temperature dependence of piezoresponse using simplified model [Eq.(30)] and rigorous calculation in the WI limit results in very similar temperature dependences. This is due to the fact that in both cases the response is determined by (almost) temperature independent ratios of the type $d_{ij}/\varepsilon_{ij} \sim PQ_{ij}$, rather than strongly temperature dependent piezoelectric coefficients. In the simplified model only $d_{33}/\varepsilon_{33}$ ratio is considered, while in the WI description all relevant parameters are incorporated. In the WI description, however, the physical origins of this weak temperature dependence are less obvious.

Thus, the temperature dependence of experimental PFM contrast suggests that under the experimental conditions ($F_0 \approx 200$ nN, nominal radius $R \approx 30$ nm, tip is not blunted) the imaging mechanism of PFM is governed by the dielectric gap effect. The major contribution to piezoresponse is an electromechanical response of the surface to the tip bias, however, the properties of tip-surface contact change with temperature. The width of the "apparent gap" in these measurements can be estimated as > 1 nm. This conclusion is verified by small experimental piezoresponse coefficients (~ 4 pm/V) [45,53,91,92] as compared to the calculated value for $BaTiO_3$ (~ 50 - 100 pm/V).

## 4.4. Simultaneous Acquisition of PFM and Potential Images

It is clear from Section 2.3 and 3.3 that the information provided by non-contact and contact electrostatic SPMs on ferroelectric surfaces is complementary. The former quantifies the field and potential related to the surface chemistry as well as local polarization, while the latter quantifies more intrinsic electromechanical response with possible contribution of electrostatic interactions. Only for well-defined surfaces the qualitative information on domain morphology obtained by these techniques coincides. Therefore, simultaneous acquisition of potential and PFM images is a matter of considerable experimental and theoretical interest. This is especially important for investigations of dynamic phenomena, in which large time intervals between sequential PFM/SSPM images are unacceptable. Under equilibrium conditions, simultaneous acquisition of piezoresponse and potential images can facilitate the correlation between topographic, potential and piezore-sponse features and analysis of surface properties. Simultaneous PFM and SSPM imaging can be implemented using the usual lift mode so that the topography and piezoresponse are acquired in contact and potential is collected on the interleave line. Figure 21 illustrates several examples of simultaneous piezoresponse and potential imaging on BaTiO$_3$ and PZT. An open loop version of SSPM is used. For BaTiO$_3$ both SSPM and PFM features are related to the surface domain structure and, therefore, are closely correlated. For PZT the information provided by the two is complementary. However, after polarization switching the regions with deposited charge and reversed polarization are distinguished. This illustrates the approach to independently obtain information that allows capacitive vs. electromechanical interactions to be quantified.

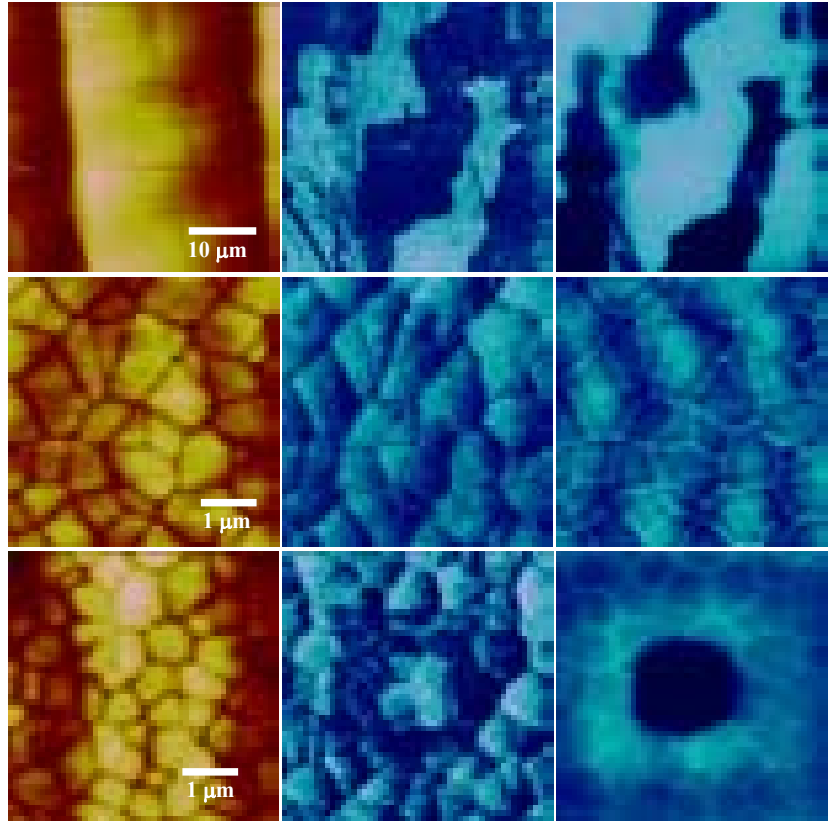

**Fig. 21.** Surface topography (left), piezoresponse (central) and open-loop SSPM (right) images from *a-c* domains on the BaTiO$_3$ (100) surface (top), for a pristine PZT surface (middle) and for PZT after switching by 10 V at 2.5 μm and -10 V at 1 μm. Potential and piezoresponse images are obtained simultaneously.

## 4.5. Conclusions

To summarize, electrostatic SPMs are powerful quantitative tools to study surface chemistry and physics of ferroelectrics. On surfaces with known crystallographic orientation the surface domain structure can be determined. Surface topography in ferroelastic materials is directly related to the angle between domains with different polarization directions, e.g. for tetragonal perovskites the corrugation angle, $\theta$, associated with 90° *a-c* domain walls is $\theta = \pi/2 - 2\arctan(a/c)$, where *a* and *c* are the parameters of the tetragonal unit cell. Complimentary information on surface potential obtained by non-contact (SSPM, EFM) or contact (PFM) SPM allows the orientation of polarization vectors (e.g. $c^+$ - $c^-$ domains) to be distinguished, thus providing a reconstruction of surface domain structure.

Combination of SSPM and EFM data indicates that the $BaTiO_3$ (100) surface is completely screened in air. At room temperature surface potential has the sign of the screening charges and is reverse to that expected from polarization orientation, i.e. $c^+$ domains are negative and $c^-$ domains are positive on the SSPM image. Temperature and time dependent behavior of surface potential is governed by rapid polarization dynamics and slow screening charge dynamics. Increasing the temperature results in a decrease of polarization bound charge leaving the screening charges uncompensated, thus increasing the effective surface potential. On decreasing the temperature spontaneous polarization increases and for a short period of time the sign of domain potential is determined by the polarization charge, thus giving rise to temperature induced domain potential inversion. Under isothermal conditions, polarization and screening charges equilibrate and the potential achieves an equilibrium value. The relevant thermodynamic and kinetic parameters can be obtained from the SPM data, which thus provide a powerful tool for the investigation of spatially constrained chemical reactions.

Response mechanisms in PFM are complex and include electrostatic, electromechanical and non-local contributions. Analytical models for electrostatic and electromechanical contrast in PFM have been developed. Image charge calculations are used to determine potential and field distributions in the tip-surface junction between a spherical tip and an anisotropic dielectric half plane. For high dielectric constant materials the surface potential directly below the tip is significantly smaller than the tip potential, implying the presence of an effective dielectric gap. The effect of the unscreened polarization charge during PFM is estimated and is shown to be negligible under ambient conditions for $BaTiO_3$. Within the electromechanical regime, strong (classical) and weak (field induced) indentation limits were distinguished. These solutions can be extended to domains of random orientation and to the analysis of stress effects in thin films by using renormalized effective electromechanical constants. Expressions for potential and field in the tip-surface junction and in the ferroelectric provide a framework for analyzing polarization switching phenomena and quantification of local hysteresis loops. The contributions of different electroelastic constants of the material to response amplitude were investigated and an almost linear correlation between piezoresponse and $d_{33}$ was illustrated for a series of PZT materials in the strong in-dentation regime. These solutions are represented by Contrast Mechanism Maps that elucidate the effect of experimental conditions on PFM. Based on these solutions the temperature dependence of piezoresponse on a $BaTiO_3$ surface was interpreted in terms of weak indentation/dielectric gap model, resolving apparent inconsistency between the divergence of $d_{33}$ at the Curie temperature and the experimental decay of the PFM signal with temperature.

Simple quantitative criterion for non-local cantilever-surface interactions in PFM is developed. The effective displacement due to cantilever buckling is inversely proportional to the spring constant of the cantilever. Depending on cantilever geometry, non-local interactions are small for cantilevers with spring constants $k_{eff} > 1$ N/m. This analysis can be used to introduce a non-local cantilever correction to local hysteresis loops obtained by PFM.

An approach for simultaneous acquisition of piezoresponse and surface potential image was developed. These data were shown to be complementary for the general case.

## Acknowledgements

Authors acknowledge the financial support from NSF Grant DMR 00-79909 and DMR 00-80863 and DoE grant DE-FG02-00ER45813-A000DOE. Discussions with A.L. Gruverman (NCSU), A.E. Giannakopoulos and S. Suresh (MIT), M. Cohen (UPenn) and M. Kachanov (Tufts) are greatly appreciated.